\documentclass[aps,prb,twocolumn,showpacs,amsmath,amssymb,floatfix,superscriptaddress]{revtex4-1}
\usepackage{graphicx}
\usepackage{subfigure}
\usepackage{color}
\usepackage{hyperref} 
\usepackage[version=3]{mhchem}
\usepackage{booktabs}

\newcommand {\be}{\begin{equation}}
\newcommand {\ee}{\end{equation}}

\AtBeginDocument{
\heavyrulewidth=.08em
\lightrulewidth=.05em
\cmidrulewidth=.03em
\belowrulesep=.65ex
\belowbottomsep=0pt
\aboverulesep=.4ex
\abovetopsep=0pt
\cmidrulesep=\doublerulesep
\cmidrulekern=.5em
\defaultaddspace=.5em
}

\begin{document}

\title{Wavelet imaging of transient energy localization in nonlinear systems at thermal equilibrium:
the case study of NaI crystals at high temperature}
\date{\today}

\author{Annise Rivi\`ere}
\affiliation{Universit\'e d'Orl\'eans, Centre de Biophysique Mol\'eculaire (CBM), CNRS UPR4301,
Rue C. Sadron, 45071 OrlŽans}

\author{Stefano Lepri}
\affiliation{Istituto dei Sistemi Complessi, Consiglio Nazionale
delle Ricerche, via Madonna del Piano 10, I-50019 Sesto Fiorentino, Italy}

\author{Daniele Colognesi}
\affiliation{Istituto dei Sistemi Complessi, Consiglio Nazionale
delle Ricerche, via Madonna del Piano 10, I-50019 Sesto Fiorentino, Italy}

\author{Francesco Piazza}
\email{Francesco.Piazza@cnrs-orleans.fr}
\affiliation{Universit\'e d'Orl\'eans, Centre de Biophysique Mol\'eculaire (CBM), CNRS UPR4301,
Rue C. Sadron, 45071 OrlŽans}

%
\begin{abstract}
\noindent In this paper we introduce a method to resolve transient excitations 
in time-frequency space from molecular dynamics simulations. 
Our technique is based on continuous wavelet transform of 
velocity time series coupled to a threshold-dependent filtering procedure to isolate 
excitation events from background noise in a given spectral region. By following in 
time the center of mass of the reference frequency interval, the data can be easily 
exploited to investigate the statistics of the burst excitation dynamics, by computing, for instance, 
the distribution of the burst lifetimes, excitation times, amplitudes and energies.
As an illustration of our method, we investigate transient excitations in the 
gap of NaI crystals at thermal equilibrium at different temperatures. Our results reveal
complex ensembles of transient nonlinear bursts in the gap, whose
lifetime and excitation rate increase with temperature. \\
\indent The method described in this paper is a powerful tool to investigate transient excitations in many-body 
systems at thermal equilibrium. Our procedure gives access to both the equilibrium and the 
kinetics of transient excitation processes, allowing one in principle to reconstruct the
full picture of the dynamical process under examination.
\end{abstract}

%
%
\maketitle
%
\section{Introduction}
%
\noindent   Hamiltonian many-body systems with nonlinear interactions admit quite generally 
a special class of periodic orbits, whose amplitude-dependent frequency does not resonate by construction with 
any of the linear (normal) modes (NM) and whose oscillation pattern is typically exponentially
localized in space. These modes, termed  discrete breathers (DB)~\cite{Marin:1998gj,Flach:2008aa,MacKay:1994aa}
or intrinsic localized modes (ILM),~\cite{Sievers:1988or} have been shown theoretically to exist
at zero temperature in a wide range of systems, including model lattices of beads and springs, 
such as the celebrated Fermi-Pasta-Ulam (FPU) chain,~\cite{Ford:1992aa} real 2D and 3D crystals,~\cite{Dmitriev:2015aa} 
both in the gap~\cite{Kiselev:1997lr} and above the phonon spectrum,~\cite{Hizhnyakov:2016ab} 
including cuprate high $T_c$ superconductors,~\cite{Marin:2001fj} boron nitride,~\cite{Barani:2017aa}
graphene~\cite{Evazzade:2018aa,Fraile:2016aa,Hizhnyakov:2016aa} and diamond,~\cite{Murzaev:2017aa}
disordered media~\cite{Sepulchre:1998sp,Kopidakis:1999fk,Kopidakis:2000lr} and
biomolecules,~\cite{breath-macromol} including proteins.~\cite{Juanico:2007ng,Piazza:2008mb}
Nonlinear modes of this kind are surmised to play a subtle role in many condensed-matter systems.
For example, DBs have been found to be connected to negative-temperature states 
({\em i.e.}, states for which the  derivative of entropy versus energy 
is negative) in the discrete nonlinear Schr\"odinger equation,~\cite{Iubini:2013aa}
which is relevant to the physics of Bose-Einstein condensates in optical lattices and arrays 
of optical waveguides.  
ILMs have also been surmised to accelerate the kinetics of defect 
annealing in solids~\cite{Dubinko:2015aa} and more generally to speed up heterogeneous catalysis 
processes.~\cite{Dubinko:2016aa,Dubinko:2014aa}\\
%
%
\indent If zero-temperature nonlinear excitations are well-established and fairly 
understood physical objects, when it comes to systems at thermal equilibrium the scenario proves
far more complex and thorny.~\cite{Ivanchenko:2004aa} Numerical techniques based on spectral analyses coupled to 
surface cooling techniques have been proposed as means to detect spontaneous DB excitation 
in model nonlinear lattices.~\cite{Eleftheriou:2005aa} More recently, other studies have also addressed  
this problem via equilibrium MD simulations, both in model nonlinear chains~\cite{Ming:2017aa} 
and in crystals with realistic potentials ranging from graphane~\cite{Baimova:2017aa,Baimova:2016aa}
to crystals with the NaCl structure.~\cite{Kistanov:2012aa}\\
\indent Experimental evidence for nonlinear localized excitations is no less a spinous matter. 
Nonlinear localized modes have been found experimentally at finite temperature 
in Josephson ladders~\cite{Binder:2000ih} and arrays.~\cite{Trias:2000ej}
However, the oldest experimental evidence explained in terms 
of excitation of ILMs at finite temperature in a crystal are the elusive tracks arising from nuclear scattering events 
in muscovite mica.~\cite{Steeds:1993aa} Such dark lines, known since a long time,~\cite{Russell:2015aa} have 
led to the suggestion that ILMs might act as energy carriers in crystals along specific 
directions with minimal lateral spreading and over long distances.~\cite{Marin:1998aa}
Recently, experimental evidence has been collected in support of this inference, 
as infinite charge mobility has been measured at room temperature in muscovite mica crystals
irradiated with high-energy alpha particles.~\cite{Russell:2017aa}\\
\indent Indirect evidence for the non-equilibrium excitation of ILMs at finite temperature has been also
gathered through inelastic x-ray and neutron scattering measurements on
$\alpha$-uranium single crystals.~\cite{Manley:2006aa,Manley:2008aa}
In particular, the authors of these studies speculate that the excitation of mobile modes, whose
properties are consistent with those of ILMs, could explain the measured anisotropy of thermal expansion and 
the deviation of heat capacity from the theoretical prediction at high temperatures.~\cite{Murzaev:2016aa}
More recently, the same authors have published experimental evidence of the excitation of
intrinsic localized modes  in the high-temperature vibrational spectrum of 
NaI crystals,~\cite{Manley:2009aa} where ILMs have been predicted to exist at $T=0$ and 
characterized by many authors.~\cite{Sievers:2013aa,Kistanov:2012aa,Khadeeva:2010aa,Nevedrov:2001aa,Kiselev:1997lr}
In 2011, the same authors published time-of-flight inelastic neutron scattering measurements 
performed on NaI single-crystals.~\cite{Manley:2011aa} Their results seemed to point at the spontaneous 
thermal excitation of ILMs, moving back and forth between the [111] and [011] orientations at intermediate
temperatures and eventually locking in along the [011] orientation above $T=636$ K.
Further inelastic neutron scattering measurements on NaI crystals 
published in 2014 found no evidence for thermally activated localized
modes.~\cite{Kempa:2014aa} Even though these measurements confirmed a very small peak 
within the gap, its intensity is so small -- the authors argue -- that it is nearly impossible to discern whether 
it is part of the inelastic background or whether it is indeed a true signature 
of a coherent scattering event.
However, in a subsequent paper~\cite{Manley:2014aa}, Manley and coworkers made it clear 
that the interpretation of the coherent scattering from NaI requires a
correction of the incoherent background from the incoherent cross
section of Na, which was not included in Ref~\onlinecite{Kempa:2014aa}. 
As the partial phonon DOS of Na displays a
stretch of reduced intensity at high temperatures in the spectral region corresponding 
to the $T=0$ gap, when
this correction is made (as in Ref.~\onlinecite{Manley:2014aa}), 
the ILM feature becomes a little more pronounced. Combining neutron scattering, 
laser flash calorimetry and accurate x-ray diffraction data, the authors 
then argued that ILM localization in NaI occurs in randomly stacked planes 
perpendicular to the (110) direction(s) with a complex temperature dependence~\cite{Manley:2014aa}.
As a result, they suggested that spontaneous localization of ILMs should be regarded
as some sort of collective phenomenon rather than the 
random excitation of point-like modes. \\
\indent To this complex scenario one should add that   
the expected relative fraction of light ions harboring a thermally excited ILM in NaI
is relatively low. As an example, the prediction made in Ref.~\onlinecite{Sievers:2013aa} 
for ILMs polarized along the [111] orientation at $T=636$ K
is about $8.3\times 10^{-4}$, which would 
make their direct observation a very hard matter. \\
\indent Taken together, the facts exposed above reveal a lively albeit rather intricate debate concerning 
the very existence of thermal ILMs in crystals and the means to possibly spotlight their presence 
and characterize them. In order to address these questions, 
in this paper we develop a robust numerical technique based on continuous wavelet analysis,
designed as a tool to pinpoint and characterize transient vibrational excitations in general in
many-body system, and illustrate it in the case of NaI crystals. 
The paper is organized as follows. In Sec.~\ref{sec:model}, we describe the MD simulation 
protocol and present our wavelet-based technique designed to pinpoint and characterize 
transient energy bursts in the time-frequency plane.  In Sec.~\ref{sec:results1},
we apply our technique to characterize transient excitation of energy in the gap of 
NaI crystals. In Sec.~\ref{sec:results2}, based on the assumption that the population 
of transient energy bursts detected in the gap may contain spontaneous excitation events of ILMs,
we address the problem of how to sieve them out of the burst population.
In Sec.~\ref{sec:conclu}, we summarize our main findings and discuss possible 
improvements and extensions of our method to detect and characterize spontaneous
excitation of ILMs at thermal equilibrium. 
%
\section{Simulations and wavelet analysis}
\label{sec:model}
%
\noindent In order to illustrate our approach, we have used the molecular simulation (MD) 
engine LAMMPS~\cite{Plimpton:1995aa} to simulate the equilibrium dynamics of 
a NaI crystal as a function of temperature. The simulation box comprises $N_c^3$ cubic units 
cells with periodic boundary conditions (PBC) along the three Cartesian 
directions,
each cell containing 4 Na$^+$ and 4 I$^-$ ions. For all simulations reported here, 
we have taken $N_c=10$, so that the total number of ions is $8000$.~\footnote{We observe that PBCs 
with $N_c=10$ appears a safe choice to inspect energy localization on length scales of 
the order of half/one unit cell.} 
The choice of interatomic potentials is crucial. In order to determine the best available choice, we have
scrutinized a large body of specialized literature,~\cite{Ruffa:1963aa,Rapp:1973aa,Sangster:1976aa,
Jain:1976aa,Sangster:1978aa,Shanker:1984aa,Shanker:1984ab,Kumar:1986aa,Sherry:1991aa,Sumil:2012aa}
which led us to reconstruct a total potential energy of the form 
\begin{equation}
\label{e:epottot}
\begin{split}
U(\{\mathbf{r},\mathbf{R}\}) =  \sum_{i>j} V_{++}(|\mathbf{R}_i-\mathbf{R}_j|) \\
                              + \sum_{i>j} V_{--}(|\mathbf{r}_i-\mathbf{r}_j|) \\
                              + \sum_{i,j} V_{+-}(|\mathbf{R}_i-\mathbf{r}_j|) 
\end{split}
\end{equation}
where $\mathbf{R}_i$ and $\mathbf{r}_i$ denote the position vectors of Na$^+$ and I$^-$
ions, respectively. Each pairwise contribution comprises three terms,  
\begin{equation}
\label{e:epot2}
V_{\pm\pm}(r) = \frac{Q_\pm Q_\pm}{4\pi\epsilon_0r} + W^{LR}_{\pm\pm}(r) + P^{SR}_{\pm\pm}(r) 
\end{equation}
The Coulomb energy has been computed via the Ewald method.~\cite{Ewald:1921aa} Instead of specifying a
cutoff wavevector for the Ewald sums, we have set the relative error in the calculation of 
electrostatic forces to be less than $10^{-5}$ at any given time. We have verified that our results 
did not change by requiring a more accurate estimation. 
The potential energy $W^{LR}$ accounts for a long-range potential of the (6,8) kind, namely
\begin{equation}
\label{e:epotW}
W^{LR}_{\pm\pm}(r) = -\frac{C_\pm}{r^6} -\frac{D_\pm}{r^8} 
\end{equation}
corresponding to induced dipole-induced dipole interactions ($C_\pm$)
and induced dipole-induced quadrupole interactions ($D_\pm$) computed 
via the  Kirkwood-Muller methods, i.e. using experimental measurements 
of the ionic polarizability and molar susceptibility.~\cite{Muller:1936aa,Kirkwood:1932aa}
The short-range term is well described by a Buckingham-type 
potential~\cite{Harding:1991aa} of the form
\begin{equation}
\label{e:epotP}
P^{SR}_{\pm\pm}(r) = A_{\pm\pm} \exp (-r/\rho_{\pm\pm}) 
\end{equation}
restricted to the nearest-neighbor shell (5\,\AA \ cutoff).  The values of the 
parameters in Eqs.~\eqref{e:epotW} and~\eqref{e:epotP} are listed in Table~\ref{t:param}.\\
%
\begin{figure}[t!]
\centering
\includegraphics[width=\columnwidth]{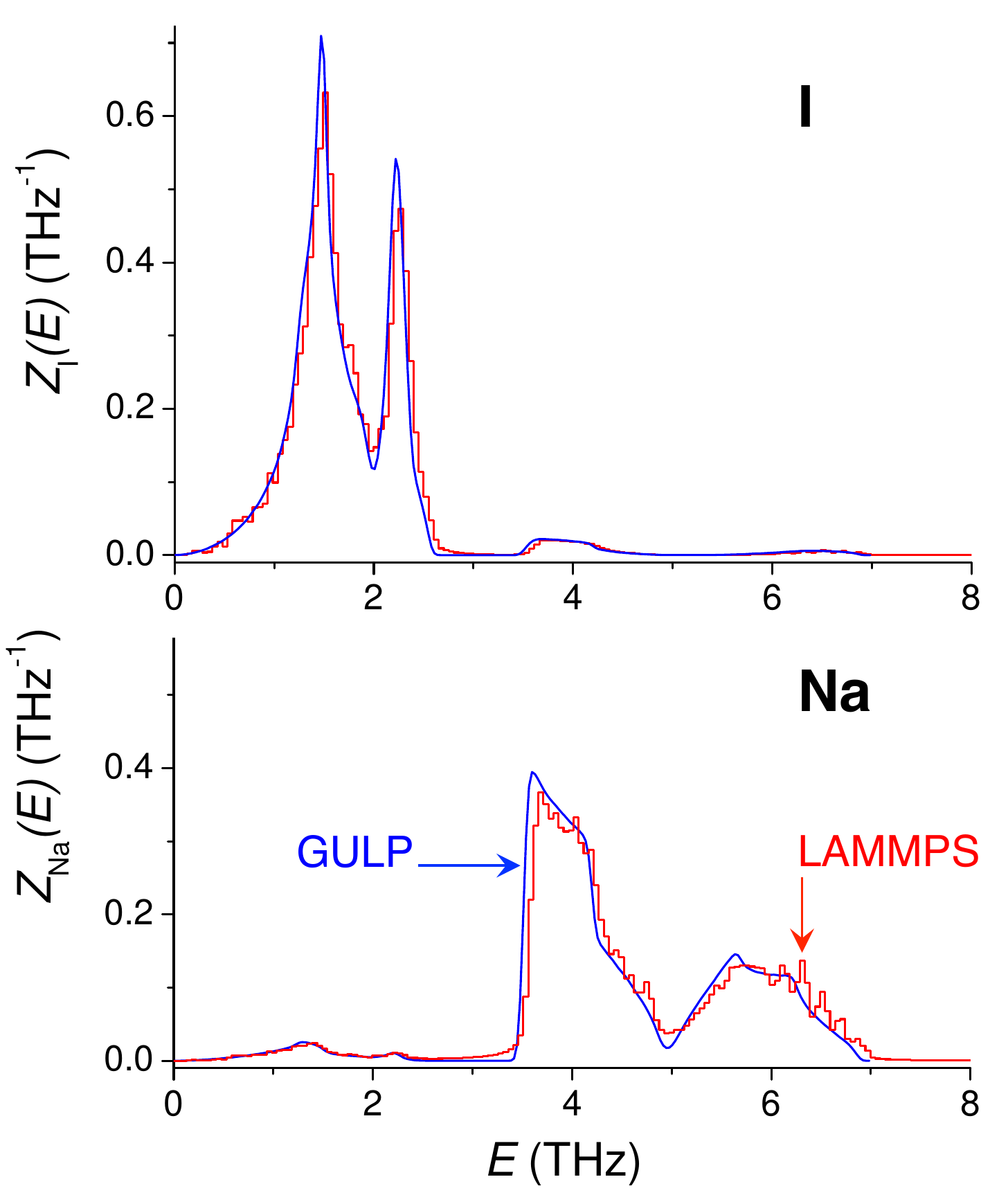}
\caption{(Color online) Phonon DOS of NaI computed from NVE MD simulations at 
($T = 38$ K, LAMMPS, red staircases) and from lattice dynamics calculations 
($T = 77$ K, GULP, blue lines). The energy $E = h\nu$ is measured in units of the frequency 
$\nu$.\label{f:DOPS}}
\end{figure}
%
%
\begin{table}[t!]
\caption{\label{t:param} 
Parameters of the pair-wise short-range and long-range 
potential energies used in this study to simulate the dynamics a NaI crystal.
For more information, see References~\onlinecite{Ruffa:1963aa,Rapp:1973aa,Sangster:1976aa,
Jain:1976aa,Sangster:1978aa,Shanker:1984aa,Shanker:1984ab,Kumar:1986aa,Sherry:1991aa,Sumil:2012aa}.}
\begin{ruledtabular}
\begin{tabular}{ccccc}
& \multicolumn{2}{c}{\em Short-range} & \multicolumn{2}{c}{\em Long-range} \\
\cmidrule{2-3}\cmidrule{4-5}
%
%
Pair kind                   &   $A_{\pm,\pm}$ [eV]  &  $\rho_{\pm\pm}$ [\AA] &
$C_{\pm,\pm}$ [eV\AA$^6$]   &   $D_{\pm,\pm}$ [eV\AA$^8$] \\ 
\cmidrule{1-1}\cmidrule{2-3}\cmidrule{4-5}
$++$   &  8500.74                     &   0.29333
       &  4.93337                     &   3.55827                        \\
$--$   &  384.924                     &   0.50867 
       &  810.714                     &   805.769                        \\
$+-$   &  736.498                     &   0.40100
       &  54.9164                     &   47.0954                        \\
\end{tabular}
\end{ruledtabular}
\end{table}
%
%
\begin{figure*}[t!]
\centering
\includegraphics[width=15truecm]{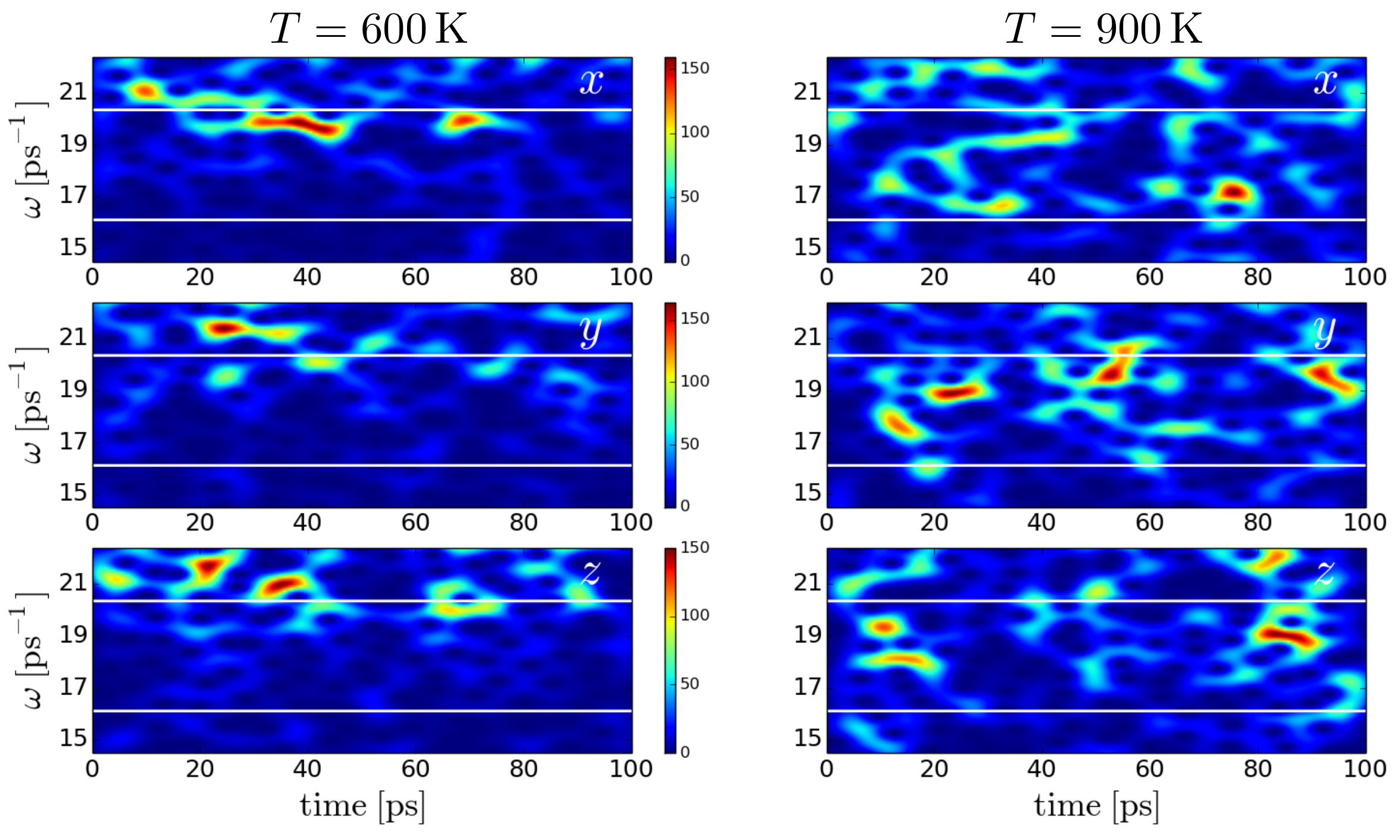}
\caption{(Color online) Time-frequency density maps of the function $|G_{i\alpha}(\omega,t)|^2$
in the gap region  for two different Na ions at $T=600$ K and $T=900$ K 
along the three Cartesian directions ($x,y,z$ from top to bottom). Spectral power is
color-coded from blue (low energy) to red (high energy).
The two horizontal white lines mark the edges $\omega_1,\omega_2$ of the gap region.
\label{f:maps}}
\end{figure*}
%
\indent Since the lattice constant of NaI crystals is known experimentally and has been used,
alongside other experimentally determined constants, to parametrize the potential 
energy~\eqref{e:epottot},~\cite{Ruffa:1963aa,Rapp:1973aa,Sangster:1976aa,
Jain:1976aa,Sangster:1978aa,Shanker:1984aa,Shanker:1984ab,Kumar:1986aa,Sherry:1991aa,Sumil:2012aa}
we have used these measurements to set the dimension of the unit cell at different temperatures 
and performed fixed-volume simulations.   
A typical simulation consisted of a first thermalization NVT stage of duration $\Delta t_{\rm th}$, 
where the system was brought to thermal equilibrium through a Nos\'e-Hoover 
thermostat~\cite{Nose:1984aa,Hoover:1985aa} starting from zero initial atomic displacements
and random velocities drawn from a Maxwell distribution. We have verified that 
$\Delta t_{\rm th} = 5$ ps was sufficient to correctly thermalize our system for temperatures 
larger than 400 K. Once the system is thermalized, we run constant energy trajectories (NVE)
of duration $\Delta t_p$ for data 
production.
It is interesting to remark that distortions 
driven by the localization of nonlinear vibrational modes are expected to conserve 
volume, as it was found for the internal distortions associated with ILM localization 
in the fault-like planar structures reported in Ref.~\onlinecite{Manley:2014aa}
~\footnote{The use of an NVT dynamics for production runs does not appear 
to make sense in this study. In fact, thermostats 
are in principle nothing but {\em smart sampling} techniques, designed to 
produce time series sampled from the  canonical measure. However, there 
is absolutely no guarantee that the actual trajectories (i.e., the actual {\em dynamics}) 
make any physical sense. In particular, all 
vibrational coherences are either (artificially) damped or completely destroyed,
depending on the value of the relaxation time scale chosen for the specific thermostat. 
In practice, it is preferable to switch off the thermostat once the system has 
reached thermal equilibrium, so that no artificial noise is left to fiddle with the
vibrational coherences that might emerge in specific frequency regions.}.
The results presented in the following refer to 
$\Delta t_p = 100$ ps, which afforded a reasonable compromise between computational costs and 
solid statistics. The time step used in the MD simulations was 0.001 ps.\\
\indent Fig.~\ref{f:DOPS} illustrates the comparison of the low-temperature phonon density 
of states computed by Fourier transforming the velocity-velocity autocorrelation functions
computed from our LAMMPS NVT trajectories with the results from lattice dynamics calculations 
performed with the GULP package.~\cite{Gale:2003aa} The excellent agreement validates our 
MD simulation protocol and in particular the values of the phonon frequencies that 
define the gap at zero temperature, 
i.e. $\omega_1 = 16.104$ ps$^{-1}$ (upper edge of the acoustic band)
and $\omega_2 = 20.343$ ps$^{-1}$ (lower edge of the optical band).\\

\subsection{Wavelet imaging of transient energy bursts in the gap}

\begin{figure*}[t!]
\centering
\includegraphics[width=16 truecm]{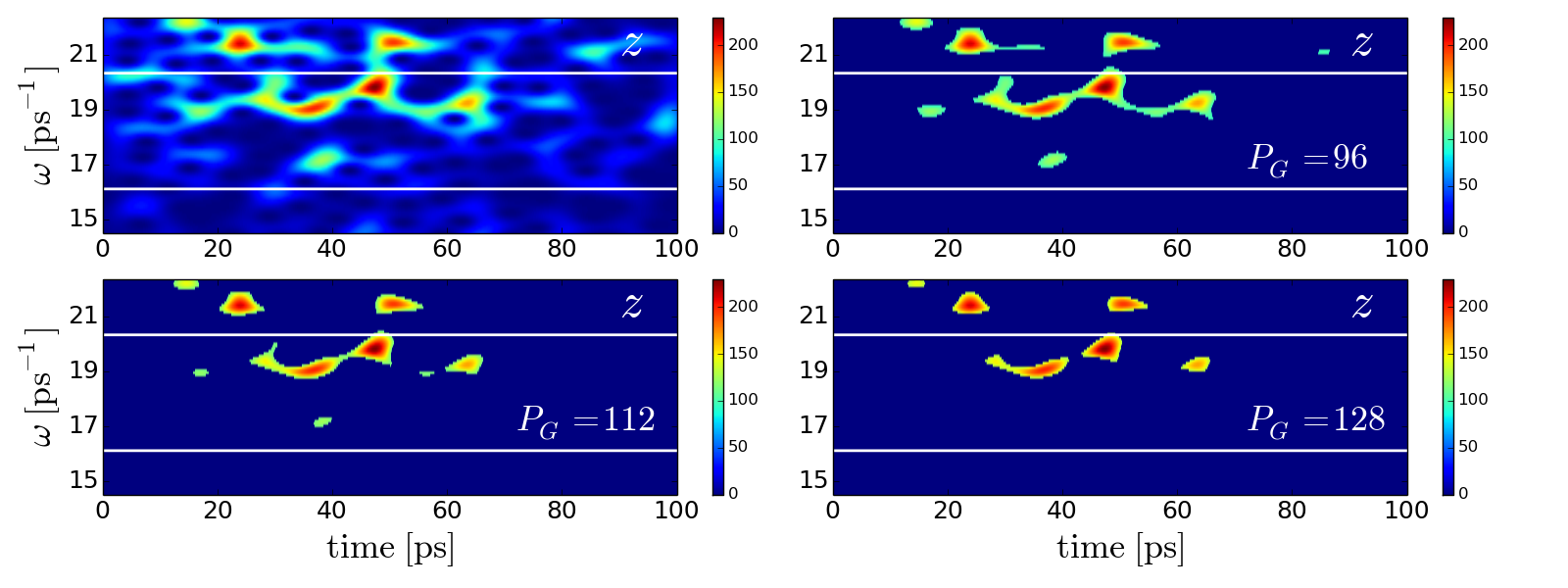}
\includegraphics[width=14 truecm]{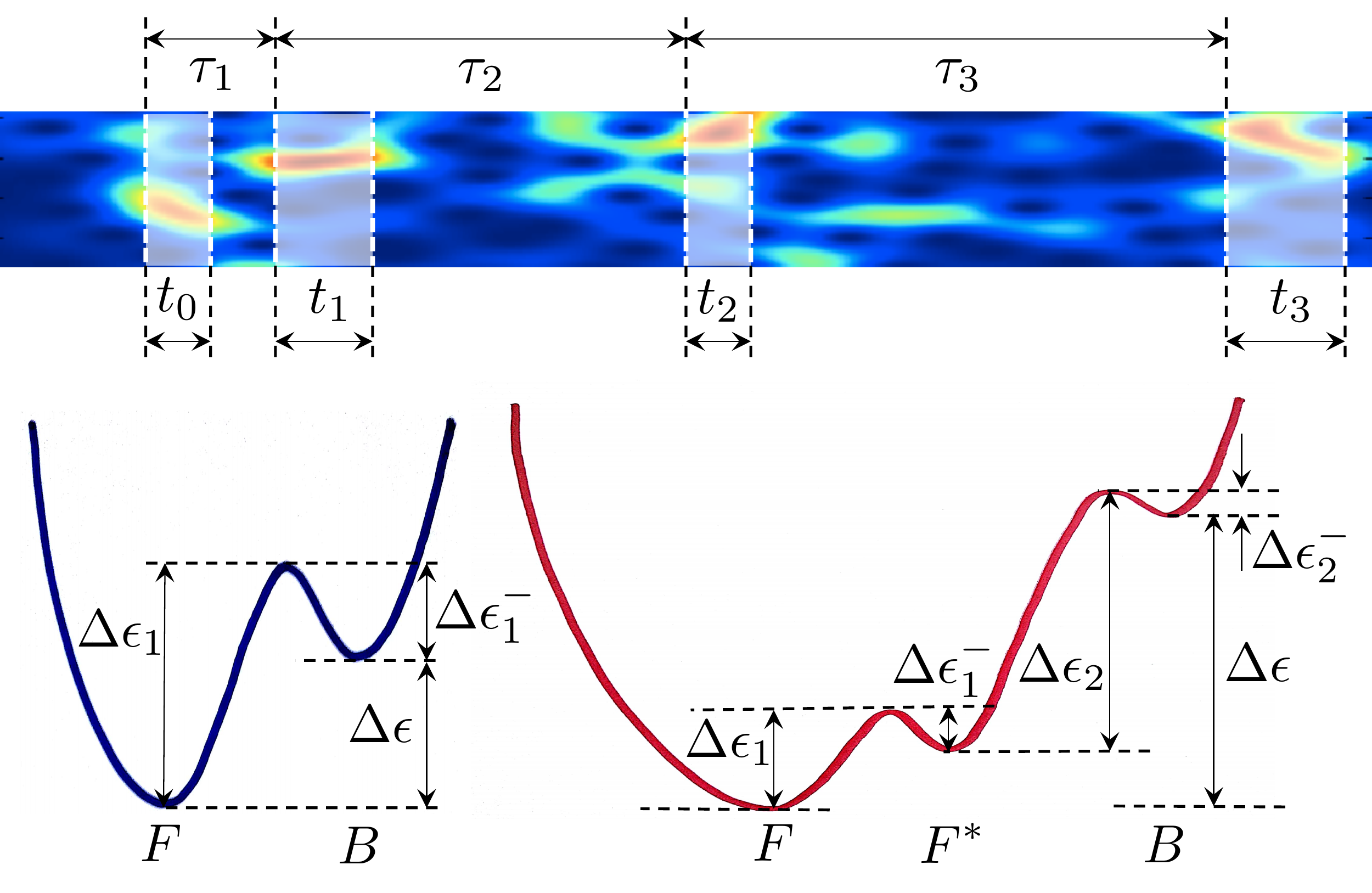}
\caption{(Color online) Top: illustration of the filtering procedure to isolate energy 
bursts with three different thresholds (units of \AA$^2$).  
Middle: scheme of the algorithm to identify lifetimes $t_n$ and 
excitation times $\tau_n$ during the production run $\Delta t_p$ for a given ion 
from the time-series of  $\langle \omega_{i\alpha}(t) \rangle$ defined in eq.~\eqref{e:gapw}.
Bottom left: kinetic model based on a two-well landscape fails to reproduce the 
kinetics and equilibrium properties of burst excitation. 
Bottom right: at least one intermediate state is required to rationalize the kinetics and 
equilibrium of the thermally activated  process of burst generation. 
This profile reproduces to scale a possible three-well landscape that is in agreement 
with our simulation data. The energy scale that controls the burst lifetimes
in this picture is 
$\delta\epsilon := (\Delta \epsilon_1 - \Delta \epsilon^-_1) - \Delta \epsilon^-_2$
(see extended discussion in the text).
\label{f:burst_analys}}
\end{figure*}

\noindent Wavelet analysis is the ideal tool to analyze non-stationary signals in the time-frequency 
domain in order to characterize transient frequency components appearing at specific times and perduring 
for finite lapses of time. As a matter of fact, Forinash and co-workers have shown 20 years ago 
that this kind of tools can provide precious information on the dynamics of discrete breathers at zero 
temperature in nonlinear chains.~\cite{Forinash:1998aa} 
Thus, it appears natural to extend this line of reasoning to explore transient 
nonlinear localization in real crystals at thermal equilibrium.   
In this work, we have computed the Gabor transform~\cite{Gabor:1946aa} of 
the time series of atomic velocities, namely
\begin{equation}
\label{e:gabor}
G_{i\alpha}(\omega,t) = \int_{-\infty}^{+\infty} e^{-(t-\tau -\Delta t_p/2)^2/a} e^{-i\omega \tau } v_{i\alpha}(\tau ) \, d\tau 
\end{equation}
where  $v_{i\alpha}$ is the velocity of the $i$-th ion along the Cartesian direction $\alpha$.
We have set the resolution parameter $a=20$ ps$^2$, optimized so as to maximize the resolution 
in both the time and frequency domains. \\
\indent As an illustration of our analysis, 
fig.~\ref{f:maps} shows typical density maps of $|G_{i\alpha}(\omega,t)|^2$
computed from the velocity time series of two random Na ions at $T=600$ K and $T=900$ K. 
It can be appreciated that, as the temperature increases, transient energy bursts 
pop up increasingly deep in the gap and persist with lifetimes of the order of 
up to 10 ps, during which their frequency appears to drift to a various degree.
In order to separate energy bursts from the background and perform a full temperature-dependent 
statistical analysis of the excitation dynamics,  it appears natural to impose a threshold $P_G$ 
on the Gabor power so as to eliminate transient background noise.
To this end, we define the filtered normalized two-dimensional 
excitation density  $\rho_{i\alpha}(\omega,t)$ as 
\begin{equation}
\label{e:rhoG}
\rho_{i\alpha}(\omega,t) = 
\frac{|\widetilde{G}_{i\alpha}(\omega,t)|^2}
     {\int_{\omega_2}^{\omega_1}|\widetilde{G}_{i\alpha}(\omega^\prime,t)|^2\,d\omega^{\prime}}
\end{equation}     
where 
\begin{equation}
\label{e:Gthresh}
\widetilde{G}_{i\alpha}(\omega,t) = 
\begin{cases}
G_{i\alpha}(\omega,t) & \text{for} \quad |G_{i\alpha}(\omega,t)|^2 \geq P_G \\
0                     & \text{otherwise}
\end{cases}
\end{equation}
This definition allows us to compute the time-dependent moments of $\rho_{i\alpha}(\omega,t)$,
which provide important information on the dynamics of transient energy excitation 
in the gap. In the present work, we concentrate on the first moment, namely
\begin{equation}
\label{e:gapw}
\langle \omega_{i\alpha}(t) \rangle = 
\int_{\omega_1}^{\omega_2} \omega \rho_{i\alpha}(\omega,t) \, d\omega
\end{equation}
As it can be seen from the top panel in Fig.~\ref{f:burst_analys}, the choice of the threshold 
$P_G$ sets the resolution limit of individual burst events. After careful examination of many 
such events, we have fixed  $P_G = 128$ \AA$^2$, 
which ensures that consecutive bursts should be optimally resolved. Although the results reported 
in the following refer to this (rather conservative) choice, we have repeated our analyses with the two 
lower values of $P_G$ shown Fig.~\ref{f:burst_analys}. While the actual figures may change slightly, 
we have verified that the relevant statistical and physical properties of the burst excitation 
dynamics are unchanged.\\
\indent After the filtering and integration procedure for a given ion $i$, 
the time series $\langle \omega_{i\alpha}(t) \rangle$ are piecewise composed of 
stretches of consecutive zeros (absence of a burst) and consecutive non-zero values,
each representing a burst and extending over its corresponding lifetime.
Such values describe the drift of the center-of-mass frequency of the burst since the moment of
its excitation until it collapses.
From the support of these time series, it is then straightforward to obtain other 
restricted time series {\em per burst}, most importantly the sequences of kinetic energies and vibration 
amplitudes for each burst during its lifetime.

%

\section{Results I -- transient energy bursts in the gap with increasing lifetimes}
\label{sec:results1}

\begin{figure*}[t!]
\centering
\includegraphics[width=17truecm,clip]{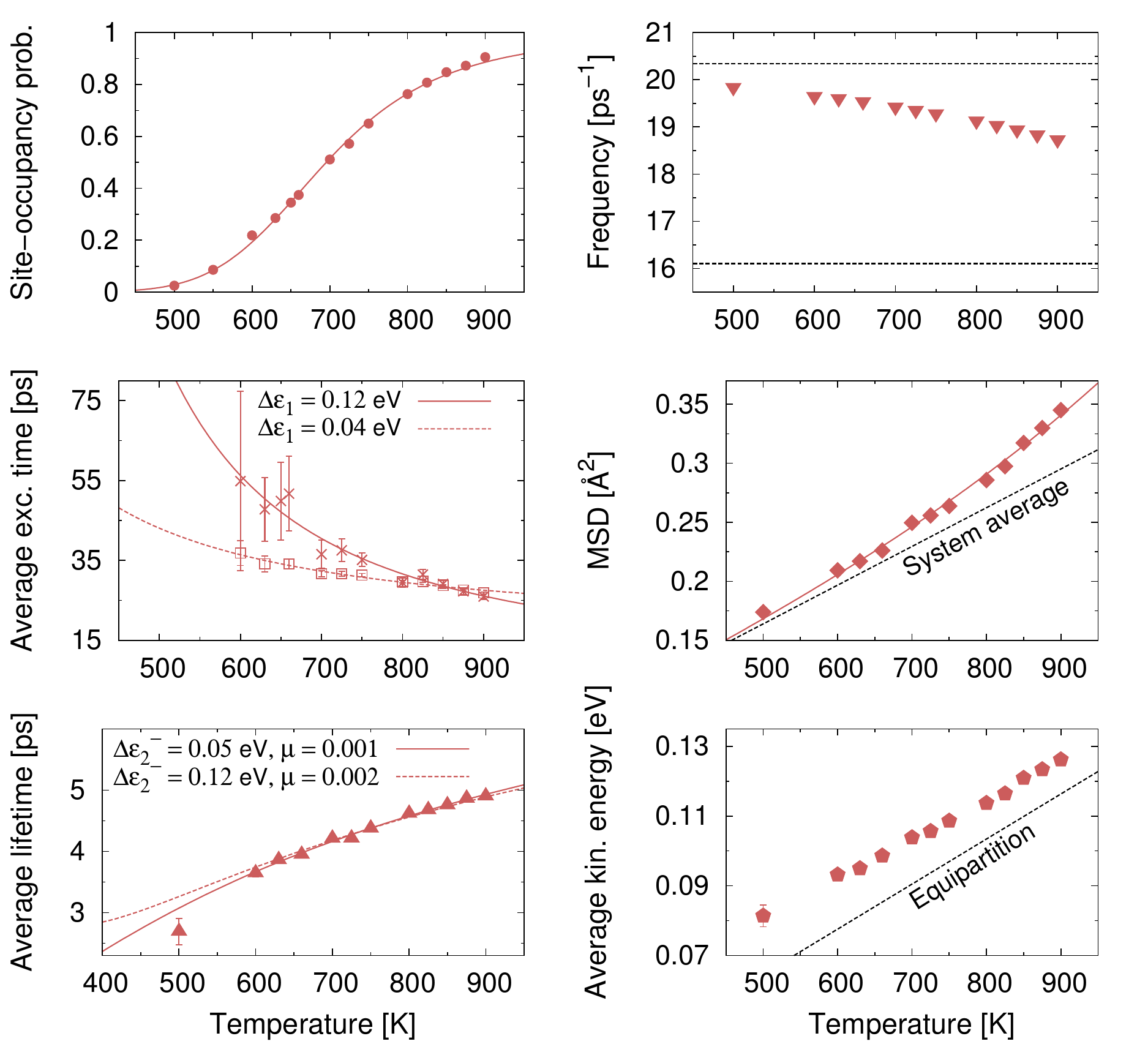}
\caption{
Analysis of the burst excitation equilibrium, kinetics and dynamics. 
Top left: equilibrium burst site-occupancy probability at Na ions vs temperature
from the simulations (filled circles) and fit with the chemical equilibrium model~\eqref{e:burstEQ}.
Best-fit parameters are:  $\Delta\epsilon = 0.54 \pm 0.01$ eV, $\Delta s = 9 \pm 0.2$ $k_B$.
Middle left: average excitation times (see again Fig.~\ref{f:burst_analys}) 
identified from the support of the filtered integrated time series~\eqref{e:gapw}.
Open squares represent the average values computed over all the pairs of 
consecutive excitation events, further averaged over $x,y$ and $z$.
The crosses represent the values computed by fitting the exponential
tails of the distributions and rescaled so as to match the high-temperature averages.
This set of data is likely to better approximate the true values at low temperatures. 
The two lines are plot best-fit Arrhenius laws of the kind~\eqref{e:Arrh1}. 
Best fit parameters are: $\Delta\epsilon_1  = 0.12 \pm 0.1$ eV, $k_1^\infty = 0.18 \pm 0.03$ ps$^{-1}$
(solid line) and $\Delta\epsilon_1  = 0.04 \pm 0.02$ eV, $k_1^\infty = 0.06 \pm 0.005$ ps$^{-1}$ (dashed
line). The {\em true} value of $\Delta\epsilon_1$ 
(i.e. the average computed over a simulation long enough to sample very long excitation times)
is expected to be in the interval $[0.04,0.12]$ eV.
%
Bottom left: average lifetimes  (see again Fig.~\ref{f:burst_analys}) 
identified from the support of the filtered integrated time series~\eqref{e:gapw} (symbols) 
and fits with the three-states model expression~\eqref{e:lifethefit}.
The solid line is a three-parameter fit, where the floating parameters 
are $t_\infty$, $\delta\epsilon:=(\Delta\epsilon_1 - \Delta\epsilon^-_1) - \Delta\epsilon^-_2$,
$\mu = k_{-1}^\infty/k_{1}^\infty$ and $\Delta\epsilon^-_2$ is kept fixed at 0.04 eV. 
The dashed line is a two-parameter fit, where $\Delta\epsilon^-_2$ is kept fixed at 0.1 eV,
while this time the energy scale that physically controls the increasing trend, $\delta\epsilon$,
is kept fixed at the previous best-fit value, i.e.  $\delta\epsilon = 0.07$ eV (see
text for the full discussion).
%
Top right: average burst frequencies vs temperature. 
%
Middle right: average burst amplitude vs temperature (filled diamonds) and 
average amplitude of the fluctuations of all Na ions in the system (dashed straight 
line). The solid line is a fit with a function of the kind $\langle A^2(T) \rangle
= \alpha T + \beta  T^4$, intended as a guide to the eye. 
%
Bottom right: average burst kinetic energy vs temperature (filled pentagons),
i.e. ensemble average of the individual burst energies. The dashed line marks the 
equilibrium value $\langle \epsilon_{\rm kin} \rangle = 3 k_BT/2$. 
At each temperature, the reported average frequencies, amplitudes and kinetic energies
represent the ensemble averages of the {\em individual} average values {\em per burst}.
The latter are computed by averaging over the individual drift of each single 
burst, as identified from the support of the corresponding filtered time series~\eqref{e:gapw}.
We remind the reader that each burst is associated with a 
single Na ion and Cartesian direction. 
\label{f:sixpanels}}
\end{figure*}

\begin{figure}[t!]
\centering
\includegraphics[width=\columnwidth,clip]{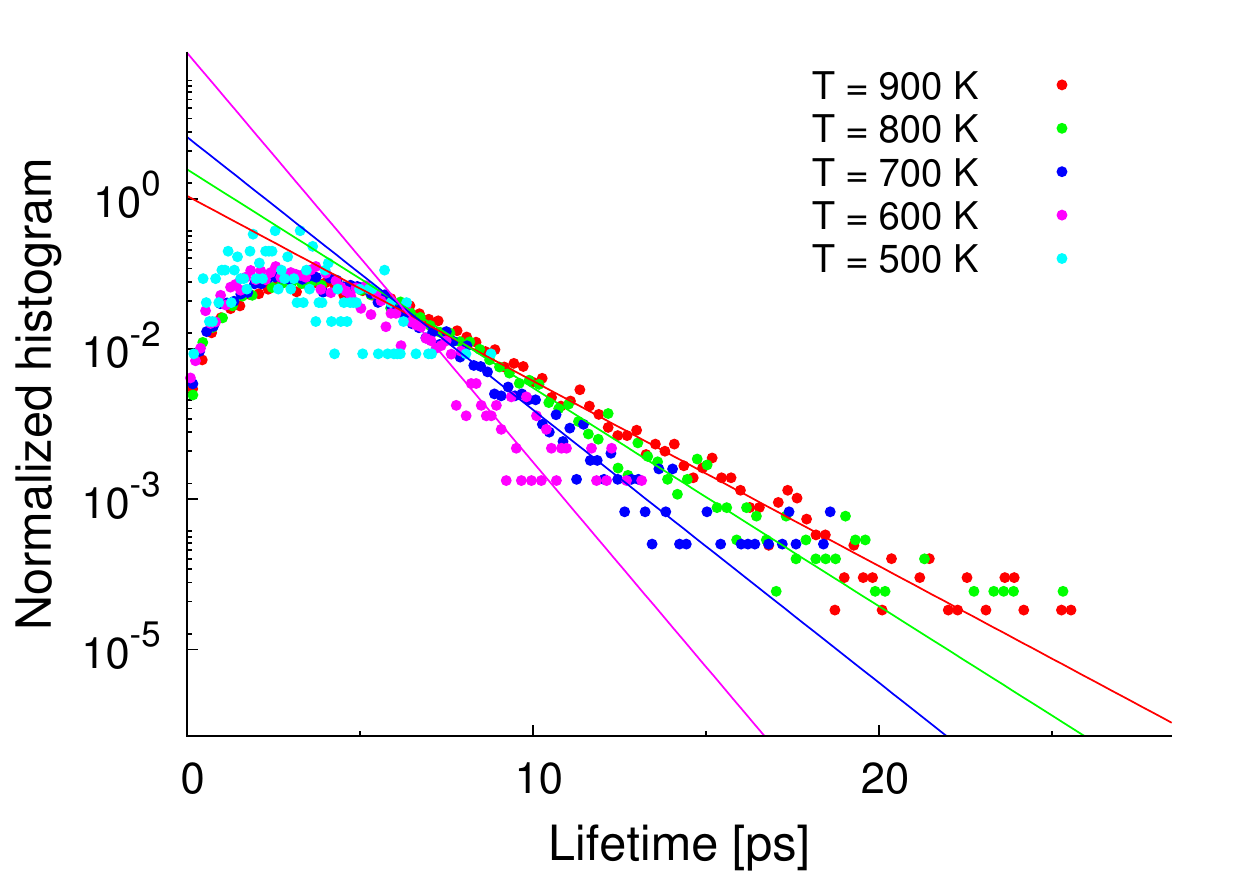}
\caption{(color online) Distributions of burst lifetimes at five representative temperatures (symbols).
The solid lines are plots of exponential fits to the distribution tails. 
\label{f:lifethis}}
\end{figure}

Nonlinear localized vibrations in the gap of diatomic lattices detach from the bottom of the 
optical band,~\cite{Cretegny:1998aa} which means that their energy is almost entirely confined 
to light ions. For a given Na ion, two key kinetics parameters describe the burst excitation dynamics, 
notably the lifetimes $t_n$ and the excitation times $\tau_{n+1}$, $n=0,1,2,\dots$. These two
measures are illustrated in the middle panel in Fig.~\ref{f:burst_analys} for a random typical 
excitation sequence. The excitation times are defined as the intervals between consecutive 
excitation events. Together with the lifetimes, they provide a rich wealth of information 
on the kinetics of burst excitation at a given temperature. However, irrespective of the 
kinetics, the temperature dependence of the site-occupancy probability (SOP) $\mathcal{P}(T)$ 
describes the equilibrium properties of this process. This can be simply computed as 
the fraction of Na ions harboring at least one burst in the gap along one of the Cartesian 
directions~\footnote{In this work we
implicitly refer to the gap spectral region when we mention the excitation of a burst.}.
The data, reported in Fig.~\ref{f:sixpanels} (top left) 
can be fitted by a simple equilibrium model of the kind 
\begin{equation}
\label{e:burstEQ}
\mathcal{P}(T) = \frac{\displaystyle 1}{\displaystyle 1 + e^{\beta \Delta f}}
\end{equation}
where $\beta = 1/k_BT$ and $\Delta f = \Delta\epsilon - T\Delta s$ is the free energy of 
burst excitation per ion. The excellent fit of the MD simulation data gives $\Delta\epsilon = 0.54 \pm 0.01$ eV
and $\Delta s = 9 \pm 0.2$ $k_B$. The data reported in Fig.~\ref{f:sixpanels} are obtained 
by averaging the site-occupancy probabilities referring to bursts along individual Cartesian directions. 
However, we observe that the three  individual SOPs  are indistinguishable from one another (data not shown),
which appears natural in view of the symmetry of the crystal.  \\
\indent It is interesting to note that the simple law~\eqref{e:burstEQ} was found to 
describe the excitation of ILMs along [111] in Ref.~\onlinecite{Sievers:2013aa}, with $\Delta\epsilon = 0.608$ eV 
and $\Delta s = 4$ $k_B$,
corresponding to the four symmetry-equivalent L points at the boundary of the Brillouin zone (BZ) from which 
an ILM can in principle detach with a [111] polarization. In our case, we  only expect a small fraction 
of the bursts to possibly be transient excitations of ILMs. It is nonetheless interesting to observe 
that the excitation energy that we find is close to a very good guess for an ILM in 3D NaI. 
Furthermore, the value $\Delta s = 9$ $k_B$ is close to the overall symmetry degeneracy of 
the L, K and X points in the BZ taken together, i.e. 10, corresponding to the extra degeneracy 
associated with the theoretical conversion points to ILMs along [110] (K) and along [100] (X).
Of course, if this interpretation has some truth to it, it seems that the three kinds of ILMs
might be excited at the same time and possibly move as units back-and-forth among them, 
as already speculated by Manley and co-workers for 
the interplay of [110] and [111] ILMs below 636 K~\cite{Manley:2011aa}.
We observe however that this kind of complex dynamics would appear exceedingly difficult 
to disentangle, even in the framework of a computational study like this,
as confirmed by the indistinguishability of the SOPs describing burst excitation 
along individual Cartesian directions.\\
\indent From the point of view of chemical kinetics, 
the expression~\eqref{e:burstEQ} describes the equilibrium between two species/states with 
an arbitrary number of intermediates. It is tempting to follow this lead to get some 
insight into the burst excitation process. In the simplest possible scenario, we 
would be dealing with two states, $F$ and $B$,
describing random  energy fluctuations ($F$) and
energy fluctuations within a burst ($B$). In the framework of this simple mean-field 
description, the time evolution of the site-occupancy probability would be given 
to a first approximation by 
\begin{equation}
\label{e:burstdyn}
\frac{\partial\mathcal{P}(T,t)}{\partial t} = 
k_1[1 - \mathcal{P}(T,t)]- k_{-1}\mathcal{P}(T,t)
\end{equation}
where $k_{1}$ and  $k_{-1}$ stand for the burst birth and death rate, respectively.
In this picture, one immediately sees that the equilibrium site-occupancy probability is
simply given by
\begin{equation}
\label{e:birtdeath} 
\mathcal{P}(T) = \frac{1}{1 + k_{-1}/k_{1}}
\end{equation}
where $k_{-1}/k_{1}$ is the effective {\em dissociation} constant of the $F-B$
equilibrium.
In a simple picture described by an energy landscape with two minima (Fig.~\ref{f:burst_analys}, 
bottom left), the excitation energy $\Delta \epsilon$ would just be the difference between the two excitation 
barriers $\Delta \epsilon_1$ ($F \to B$) and $\Delta \epsilon^-_1$ ($B \to F$), 
defined by Arrhenius-like laws of the kind 
\begin{eqnarray}
&&k_1 = k_1^\infty e^{-\beta \Delta\epsilon_1}      \label{e:Arrh1}\\
&&k_{-1} = k_{-1}^\infty e^{-\beta \Delta\epsilon^-_1}    \label{e:Arrhm1}
\end{eqnarray}
In this model $\Delta\epsilon = \Delta\epsilon_1 - \Delta\epsilon^-_1 <  \Delta\epsilon_1$ 
and  $\Delta s = \ln(k_{-1}^\infty/k_{1}^\infty)$. However, a quantitative analysis of our data 
reveals  that the best estimate of the excitation energy is 
$\Delta\epsilon_1  = 0.12 \pm 0.1$ eV, which is {\em lower} than $\Delta\epsilon$
(middle left panel in Fig.~\ref{f:sixpanels}). 
It should be stressed that the numerical determination of average excitation times is a 
delicate matter, for long excitation times are clearly under-represented in the population
of recorded events (i.e., pairs of consecutive excitations). In fact, the population observed 
in a simulation is obviously cut off at $\tau = \Delta t_p$. This means that the observed 
averages $\langle \tau(T) \rangle$ are underestimated at the lower temperatures,
where excitation times are longer. 
In order to gauge this effect, it is expedient to fit the exponential tail of the 
numerical distributions before the cutoff. The temperature trend of such decay times, 
lower in value than the corresponding  averages, should nonetheless be a good representation of 
the {\em true} trend (i.e that of averages computed from infinitely long simulations).
The middle panel in Fig.~\ref{f:sixpanels} shows that this seems, indeed, to be the case, 
placing the value of the excitation energy $\Delta\epsilon_1$ somewhere in the interval $[0.04,0.12]$ eV.\\
\indent The fact that $\Delta\epsilon_1 < \Delta\epsilon$ rules out a simple two-minima picture.
To complicate the picture further, it can be seen from Fig.~\ref{f:sixpanels} (bottom left panel)
that the average burst lifetimes are found to {\em increase} with temperature, in agreement 
with previous results of MD simulations in crystals with the NaCl structure at thermal 
equilibrium~.\cite{Khadeeva:2011aa}
As a matter of fact, we found that  the distribution of burst lifetimes extends to longer and longer times (up to 
lifetimes of the order of 20-30 ps) as the temperature increases (see Fig.~\ref{f:lifethis}). 
%
%
These somewhat counterintuitive results are also incompatible with a two-well free energy landscape, 
which would predict $\langle t(T) \rangle \propto 1/k_{-1}$ and therefore 
lifetimes decreasing with temperature, as escaping from the $B$ state becomes more 
and more favored at higher temperatures as prescribed by Eq.~\eqref{e:Arrhm1}.\\
\indent Of course, one might invoke general nonlinear effects to explain the observed 
increase in self-stabilization of bursts at increasing temperatures. However, it is not 
clear how this can be quantified in simple terms. In this paper, we explore another 
route that provides an effective description of the burst excitation dynamics and
has the advantage of sketching a general interpretative paradigm to combine 
equilibrium and kinetics observables. \\
\indent ILM/DB excitation is expected to be a thermally
activated phenomenon, in view of the general existence of excitation 
thresholds in nonlinear lattices~\cite{Flach:1997qy,Kastner:2004fk}. This has been 
confirmed explicitly for spontaneous excitation of DBs in the framework of 
surface-cooling numerical experiments  in 2D FPU lattices~\cite{Piazza:2003aa}.
If one sticks to the physics of a thermally activated process occurring along some 
reaction coordinate, in order to rationalize the observed burst excitation process,
it is necessary to introduce at least an intermediate state, $F^\ast$, 
according to the kinetic model
\begin{equation}
\label{e:FFBkin}
\ce{$F$ <=>[k_1][k_{-1}] $F^\ast$ <=>[k_2][k_{-2}] $B$}
\end{equation}
The state $F^\ast$ could be interpreted 
as a precursor fluctuation that can be either stabilized -- this is where nonlinear effects
come into play in this picture -- 
to yield a persistent burst, or it can decay back into the background. 
As we shall see in the following, 
the obvious coming into play of nonlinear effects as temperature increases is confirmed by the 
observed trend of the burst average amplitudes. The scheme~\eqref{e:FFBkin}
corresponds to a three-minima landscape as illustrated in Fig.~\ref{f:burst_analys} (bottom right panel).
The relative equilibrium population of the $B$ state, i.e., the burst site-occupancy probability
$\mathcal{P}$ in our analogy, can be simply computed by imposing the detailed-balance conditions
$k_1 F_e = k_{-1} F^\ast_e$ and  $k_2 F^\ast_e = k_{-2} B_e$. This yields immediately
\begin{equation}
\label{e:3wellEQ}
\mathcal{P} \equiv \frac{B_e}{F_e+F^\ast_e+B_e} = \frac{\displaystyle 1}
                                                        {\displaystyle 1 + 
                                                         \frac{k_{-2}(k_1+k_{-1})}{k_1k_2}}
\end{equation}                                                            
In this model, the burst lifetime is set by the rate $k_{-2}$. With reference 
to the landscape depicted in the bottom right panel in Fig.~\ref{f:burst_analys}, let us take 
$\langle t(T) \rangle \propto 1/k_{-2}$ and let us assume that $k_2$ and $k_{-2}$ are described
by Arrhenius expressions such as~\eqref{e:Arrh1} and~\eqref{e:Arrhm1}
[i.e., $k_2 = k_2^\infty \exp(-\beta \Delta\epsilon_2)$, $k_{-2} = k_{-2}^\infty \exp(-\beta \Delta\epsilon^-_2)$].
Then,  comparing Eqs.~\eqref{e:3wellEQ} and~\eqref{e:burstEQ}, we are led immediately to the following 
expression 
\begin{equation}
\label{e:lifethe}
\langle t(T) \rangle = t_\infty \left( 
                                  \frac{1 + \mu \, e^{\beta \Delta\Delta\epsilon_1 }}
                                       {1 + \mu} 
                                \right)
                                e^{-\beta (\Delta\Delta\epsilon_1 - \Delta\epsilon_2^-) } 
\end{equation}
where $\Delta\Delta\epsilon_1 := \Delta\epsilon_1 - \Delta\epsilon_1^-$, 
$\mu = k_{-1}^\infty/k_{1}^\infty$, and $t_\infty$ is the asymptotic, infinite-temperature
lifetime ($\propto 1/k_{-2}^\infty$) determined uniquely by the kinetic (entropic)
constants (see again the three-well landscape pictured in Fig.~\ref{f:burst_analys}).\\
\indent The function~\eqref{e:lifethe} is a monotonically decreasing function 
of temperature or features  a minimum at low temperatures and an increasing trend for 
higher temperatures depending on the relative value of the relevant kinetic and energy scales.
More precisely, an increasing portion at high temperature will be observed provided 
$(\Delta\Delta_1-\Delta\epsilon_2^-)/\Delta\epsilon_2^- > \mu$, that is, 
\begin{equation}
\label{e:cond}
\frac{\Delta\epsilon_1 - \Delta\epsilon_1^-}{\Delta\epsilon_{2}^-} > 1 + \frac{k_{-1}^\infty}{k_1^\infty}
\end{equation}
It should be observed that no bursts in the gap are observed in our simulations below 
$500$ K (see again the top left panel in Fig.~\ref{f:sixpanels}). This is consistent
with a barrier $\Delta \epsilon_1$ in the 0.1 eV ballpark (at 500 K the average kinetic energy 
per particle would yield a rate $k_1 \approx 0.1 k_1^\infty$). Thus the three-wells 
free energy landscape sketched in Fig.~\ref{f:burst_analys} should be considered as describing 
the stabilization of fluctuations for temperatures $\gtrsim 500$ K. The two barriers 
should be imagined as being vanishingly small at lower temperatures, where at most 
fluctuations might be described by a simple two-state $F-F^\ast$ equilibrium. 
This is the regime where bursts become short-lived and make only rare appearances in the
gap most likely close to the bottom of the optical band (see again the left panel 
in Fig.~\ref{f:maps}).\\
\indent We see from the condition~\eqref{e:cond} that, physically, increasing burst lifetimes 
at high temperatures arise as 
a combination of (i) slow decay kinetics of the intermediate state 
$F^\ast$, (ii) large values of the energy describing the $F-F^\ast$ equilibrium, 
$\Delta\epsilon_1 - \Delta\epsilon_1^-$, and small values of the energy barrier for the 
decay of the $B$ state, $\Delta\epsilon_2^-$. In particular, if the velocity constant 
of the $F^\ast \to F$ deexcitation  is much slower than the velocity of the first 
excitation, $F \to F^\ast$ (i.e., a large positive entropy difference in favor of the $F^\ast$ state),
then the term proportional to $\mu$ can be neglected and the burst lifetime will be 
an increasing function of  temperature over the whole 
physically meaningful temperature range, as controlled solely by the positive 
energy difference $(\Delta\epsilon_1 - \Delta\epsilon_1^-) - \Delta\epsilon_2^-$.\\
\indent From a practical standpoint, due to the short temperature 
stretch available to fit the numerical data and the functional form~\eqref{e:lifethe}, 
it is not possible to fit meaningfully all the unknown parameters in Eq.~\eqref{e:lifethe}.
However, the energy scale controlling the increasing trend is 
$\delta\epsilon := (\Delta\epsilon_1 - \Delta\epsilon_1^-) - \Delta\epsilon_2^-$. Hence, 
the agreement of this simple kinetic mean-field theory with the simulations can be 
assessed by fixing the unknown barrier $\Delta\epsilon_2^-$ and fitting
a functional form of the kind 
\begin{equation}
\label{e:lifethefit}
\langle t(T) \rangle = t_\infty \left( 
                                  \frac{e^{-\beta \delta\epsilon} + \mu \, e^{\beta  \Delta\epsilon_2^- }}
                                  {1+\mu}
                                \right)                                
\end{equation}
with $t_\infty$, $\mu$ and $\delta\epsilon$ free to float. For example, with 
$\Delta\epsilon_2^- = 0.04$ eV, we get $\mu = 0.06\pm 0.03$, $\delta\epsilon = 0.07\pm 0.03$ eV,
and $t_\infty = 10\pm 2$ ps. 
To obtain a more meaningful assessment, we repeated the fit by fixing the barrier to 
a different value, $\Delta\epsilon_2^- = 0.1$ eV and kept 
$\delta\epsilon = 0.07$ eV from the first fit. It is clear from Fig.~\ref{f:sixpanels} that 
the theory still describes the simulation data in the observed temperature range. 
In this case we get consistent values of the two floating parameters left, 
namely $\mu = 0.013\pm 0.03$ and $t_\infty = 11.5\pm 0.2$ ps.\\
%
%
\indent The top right panel in Fig.~\ref{f:sixpanels} shows the average 
frequency of bursts as a function of temperature. Of course, the lower edge 
of the phonon optical band is expected to soften, hence it is difficult to disentangle 
nonlinear phonon frequencies from possible ILM events from these average data as 
the gap gets progressively colonized by soft nonlinear phonons. In the following 
we will discuss this point further and point to a possible strategy to get more 
insight as to ILM signatures.\\
\indent At variance with the average frequencies, an analysis of the average vibrational 
amplitudes of bursts in the gap reveal a telltale sign 
of nonlinear effects. In the middle right panel in Fig.~\ref{f:sixpanels} we compare the 
mean square displacement (MSD) computed over all Na ions in the crystal with the average 
MSD of Na ions hosting a burst (i.e. the mean over the burst population 
of the average MSD of each burst, the latter being computed over its corresponding
lifetime). It is clear that, starting from temperatures of the order 500 K, bursts clearly 
vibrate with increasing amplitudes, detaching from the harmonic $\propto T$ law. This 
seems to indicate that bursts of energy in the gap are intrinsically nonlinear excitations.\\
\indent Another rather puzzling piece of information comes from the analysis of the average burst 
kinetic energies (lower right panel in Fig.~\ref{f:sixpanels}). These turn out to follow a 
linear trend, as the equipartition theorem would prescribe for each and every Na ion in the system, 
however the average energies seem to be proportional to an  {\em effective} temperature that 
is about 100 K higher than the true one (see the dashed line in the 
lower right panel of Fig.~\ref{f:sixpanels}). In other words, 
during the lifetime of a burst, the corresponding Na ion has on average systematically a
higher energy than the average Na ion in the system. This is in agreement with the behavior 
of the MSD.  If one surmises that the fraction of bursts that display characteristics typical of 
ILMs is non-negligible, a possible explanation of these effects might reside in the 
known tell-tale ability of ILMs to harvest energy from the background by absorbing 
lower-energy radiation.~\cite{Flach:2008aa,Piazza:2003aa} 
Pushing this line of reasoning further, the origin of the observed higher-than-average 
energies of bursts in the gap might reveal a sheer nonlinear self-stabilization process akin 
to the well-known ILM behavior during surface cooling~\cite{Piazza:2003aa} or 
akin to the properties of the so-called chaotic breathers.~\cite{Cretegny:1998ab,Kosevich:2000aa}
%
\begin{figure}[t!]
\begin{center}
\includegraphics[width=\columnwidth,clip]{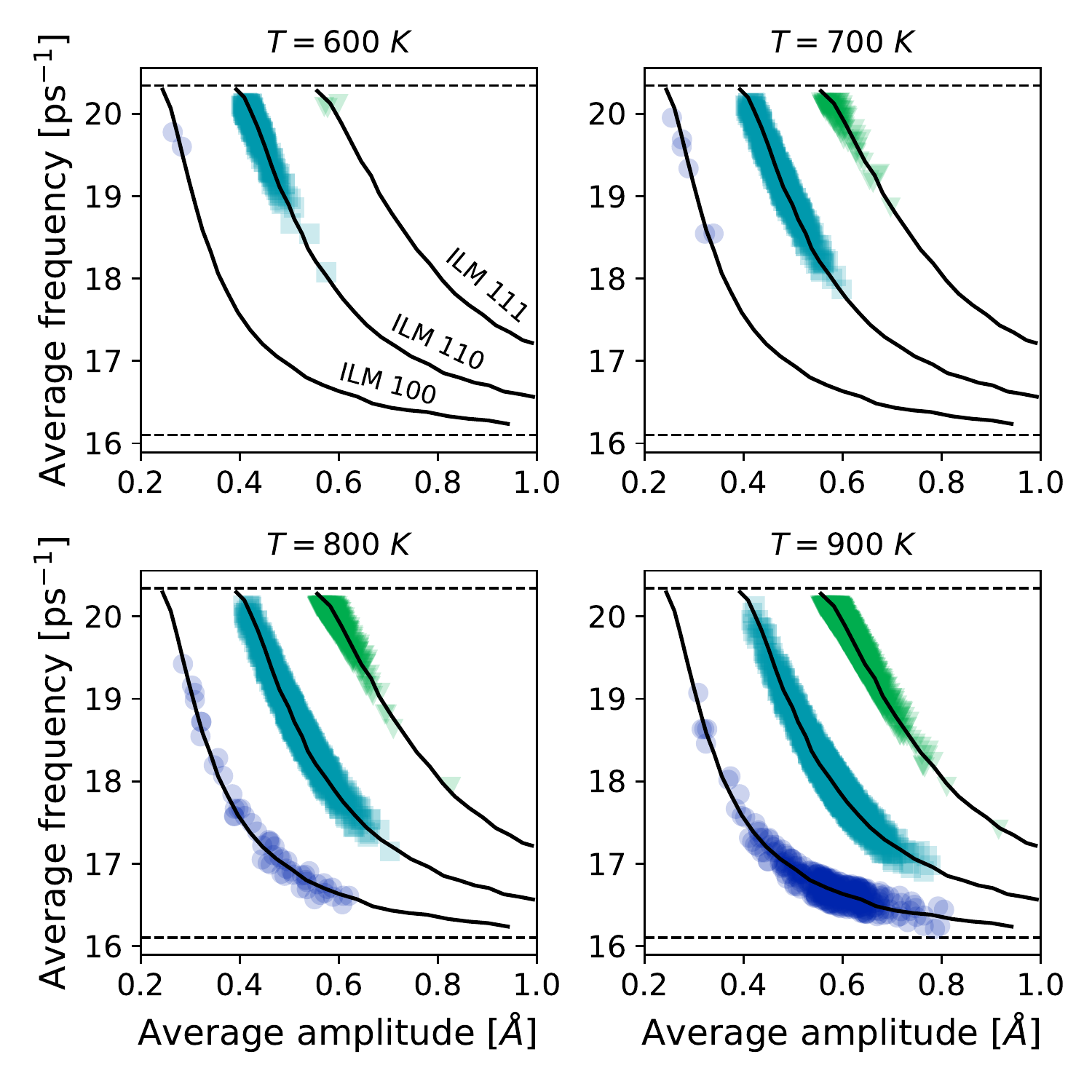}
\caption{(Color online) Illustration of the procedure employed  for sifting possible 
ILM-like excitations through the whole ensembles of bursts in the gap. At each temperature,
[100], [110] and [111] sub-ensembles are created (transparent circles)  by keeping 
only the bursts closer than 1 \% to the corresponding theoretical ILMs~\cite{Nevedrov:2001aa}
(solid lines) in the frequency-amplitude plane.}
\label{f:ILMdef}
\end{center}
\end{figure}
%
\section{Results II -- sieving through the population of bursts for ILMs}
\label{sec:results2}
%
%
\begin{figure*}[t!]
\centering
\includegraphics[width=8 truecm,clip]{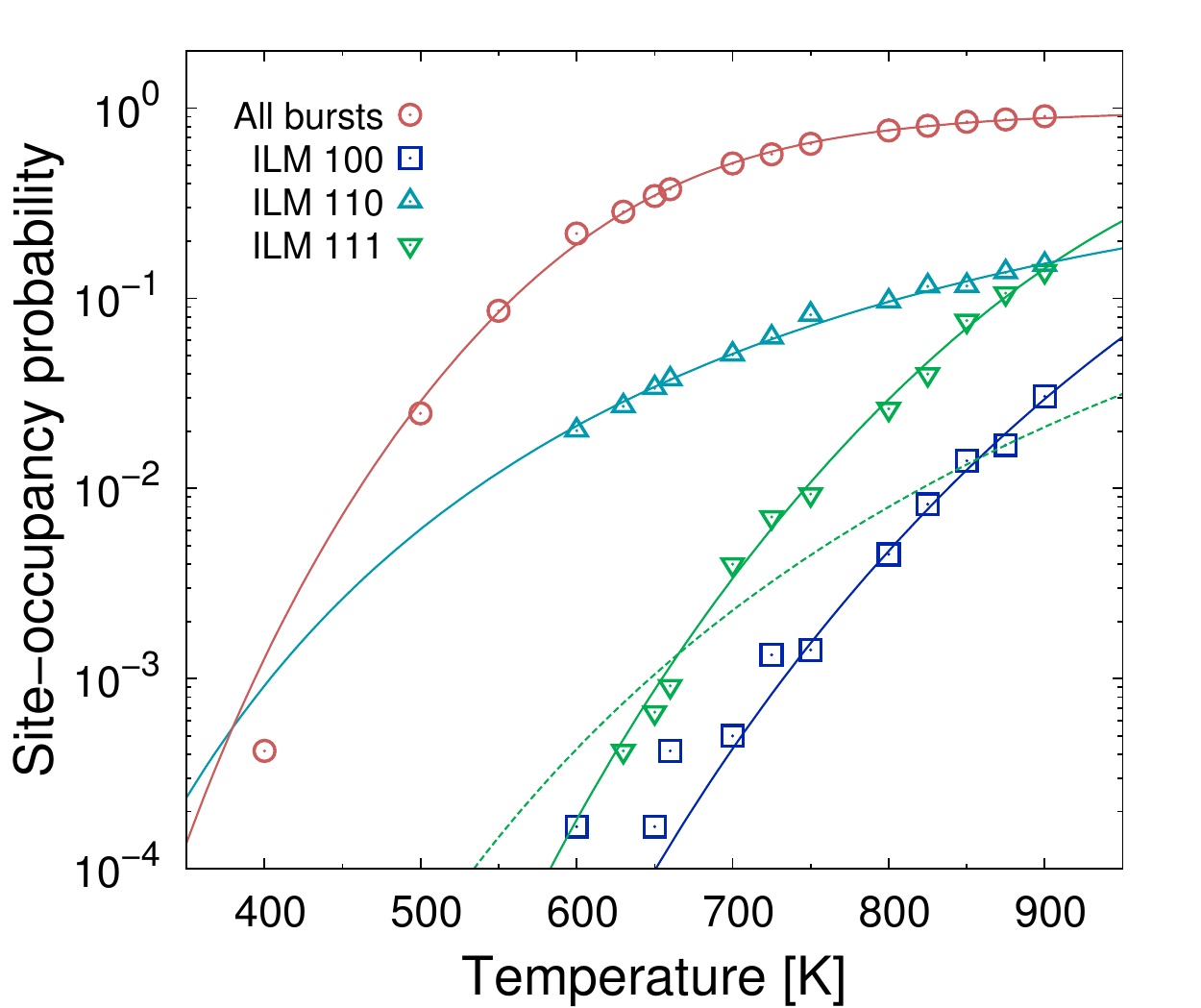}
\includegraphics[width=8 truecm,clip]{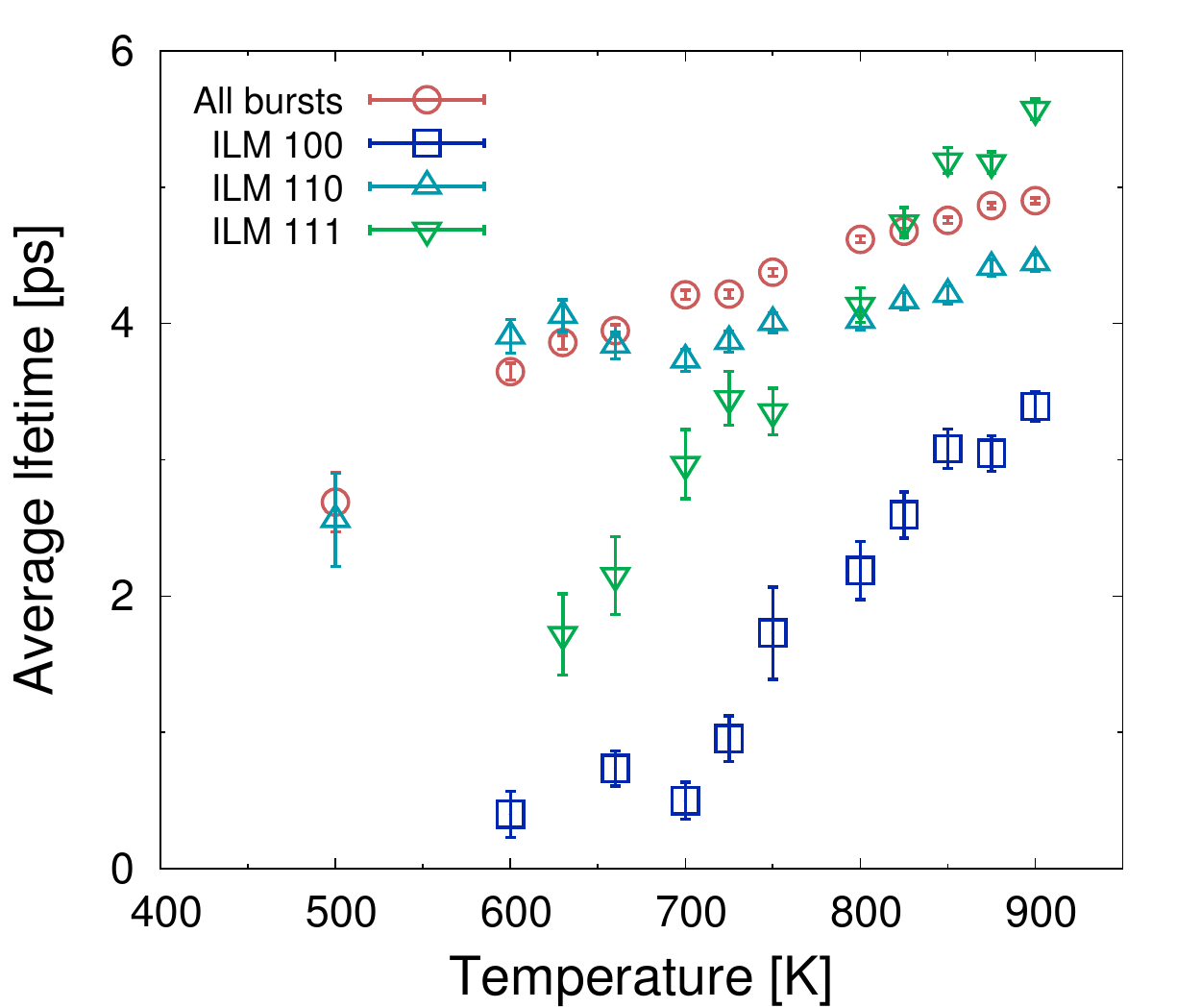}
\includegraphics[width=8 truecm,clip]{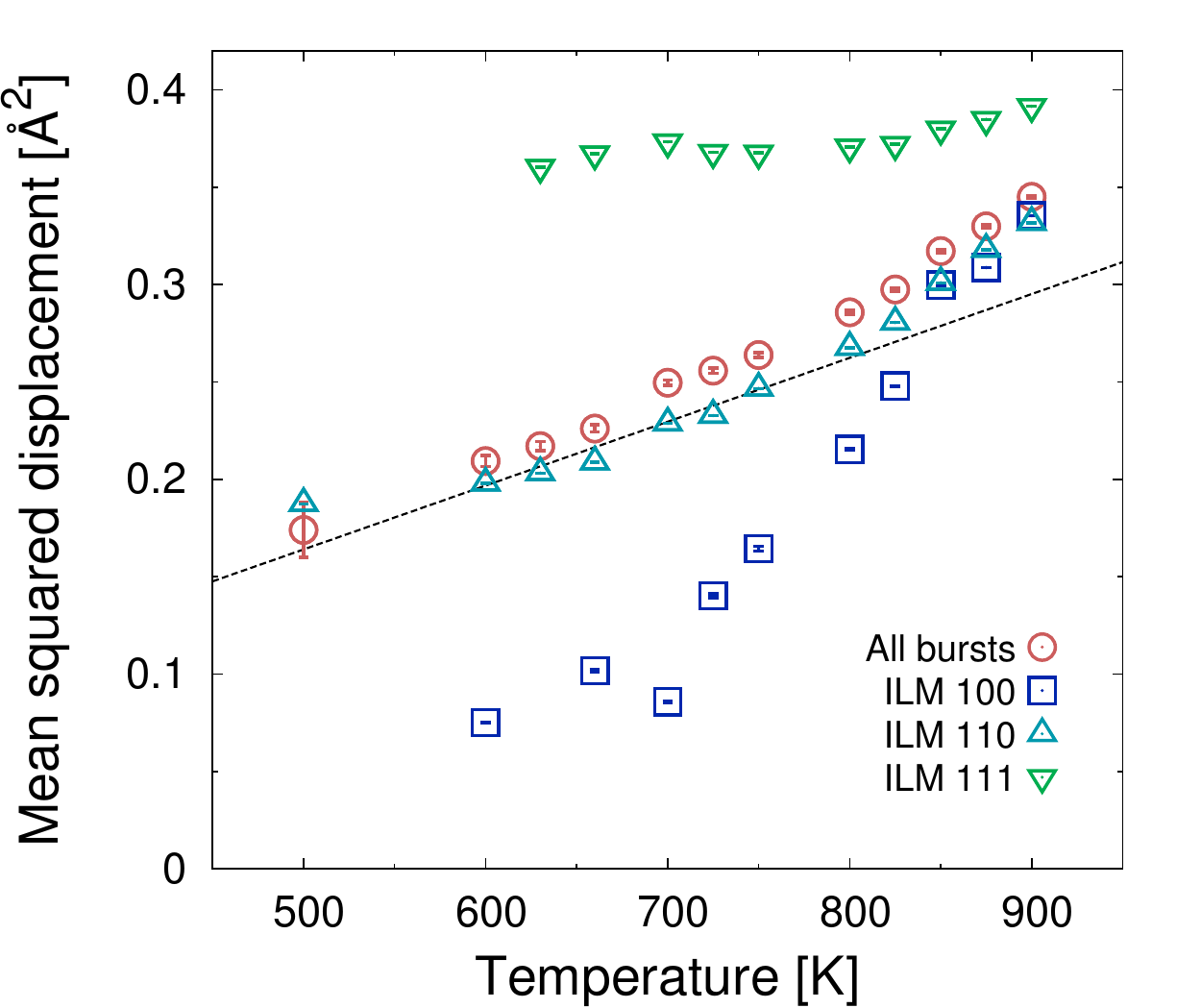}
\includegraphics[width=8 truecm,clip]{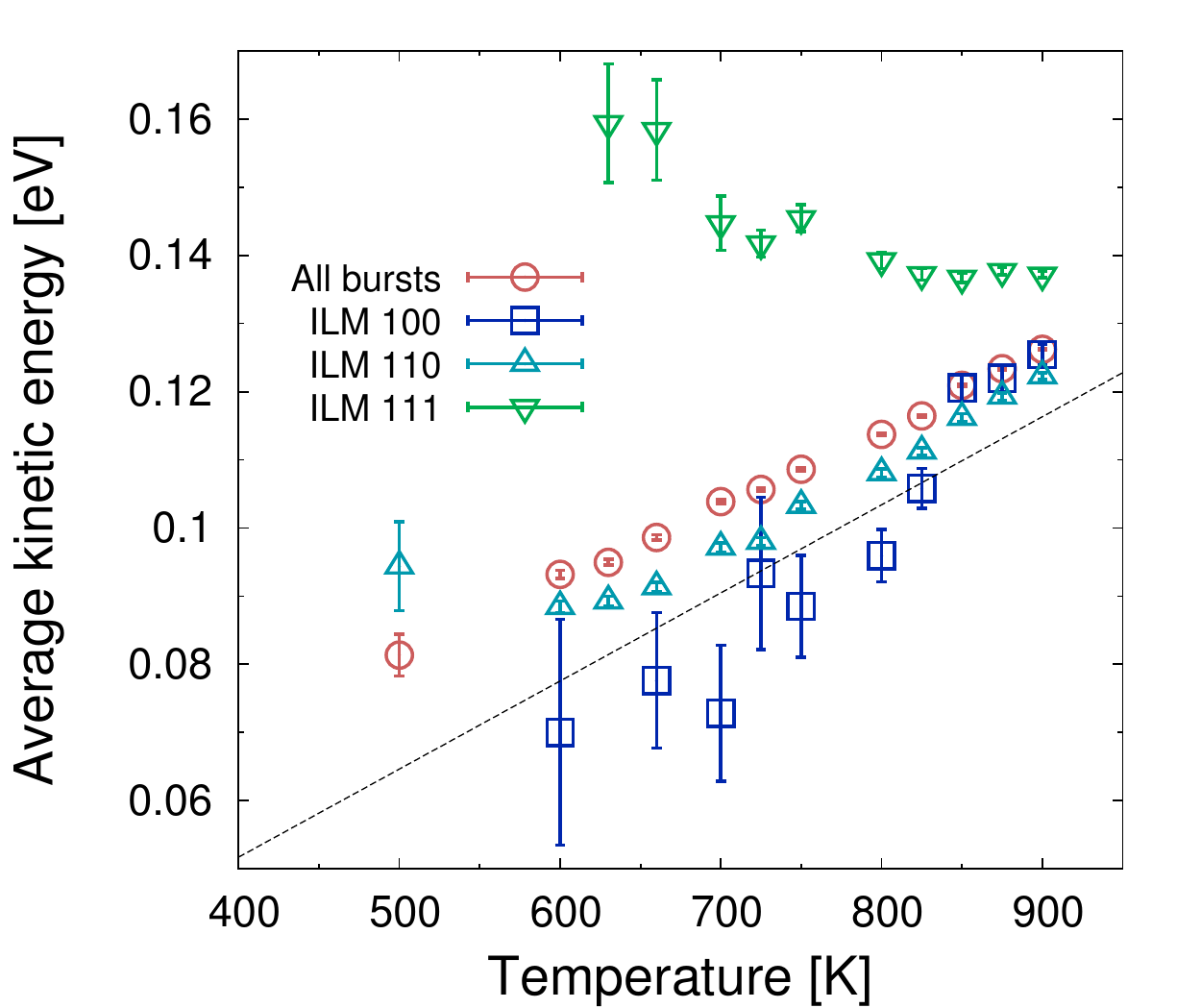}
\caption{(Color online) Burst analysis for the putative ILM sub-populations
compared to the data for the whole burst ensemble.
Top left: site-occupancy probabilities and fits with 
the expression~\eqref{e:burstEQ}. The corresponding 
best-fit parameters are reported in Table~\ref{t:ILM}. The green dashed line is the 
SOP computed in Ref.~\onlinecite{Sievers:2013aa} for [111] ILM excitations.
Top right: average lifetime. 
Bottom left: mean-square displacement. The dashed line represents the average
computed over the whole set of Na ions in the crystal. Bottom right: average kinetic 
energy. The dashed line marks the equipartition result. As expected, this 
describes the average kinetic energy of Na ions when computed over the whole 
set of Na ions in the crystal.
}
\label{f:ILM4panels}
\end{figure*}
%
\noindent The wavelet-based procedure described in this work allows one to build and characterize 
ensembles of nonlinear excitations that increasingly populate the gap as the temperature 
is raised. Even though these soft excitations display distinct ILM-like features, such as 
the apparent ability to gather some energy from the background and self-stabilize during 
their lifetime beyond the equipartition law, it is hard to state whether such bursts  are 
indeed instances of ILM excitation. In fact, according to the general arguments developed  
by Sievers and co-workers in Ref.~\onlinecite{Sievers:2013aa}, the site-occupancy probability 
of a {\em thermal} ILM is expected to be very low - about 0.02 for a $[111]$ excitation 
in 3D NaI at $T = 900$ K. While numerical analogues of exquisitely nonlinear experimental
techniques such as discussed in Ref.~\onlinecite{Wrubel:2005aa} would be powerful 
tools to address this question, it also makes sense to turn to theoretical predictions
for $T = 0$ excitations as possible {\em templates}, against which the raw 
ensembles of gap bursts can be sifted. \\
\indent The theoretical ILM frequency-amplitude relations 
reported in Ref.~\onlinecite{Nevedrov:2001aa} are shown as solid lines in Fig.~\ref{f:ILMdef}
for the three ILM polarizations, $[100]$, $[110]$, and $[111]$. At each temperature, we 
sifted through the whole collection of bursts and assembled three sub-populations 
by keeping only those excitations whose distance from the theoretical curves was less 
than 1 \%. Practically, for each burst we recovered the three theoretical frequencies corresponding to 
its measured average amplitude. The burst was then kept under the appropriate polarization 
label if the relative difference between its average frequency and the theoretical 
frequency was less than 1 \%. We observe that this is a rather crude scheme, as 
each burst is associated with a single Cartesian direction. Therefore, while 
this  procedure makes perfect sense for the $[100]$ polarization, it might be objected that 
by doing this we are not enforcing the additional correlations among different Cartesian directions
required by the assumed polarizations. Of course, a burst found along $x$ 
that would correspond to a genuine ILM polarized along the $[110]$ direction would most likely 
match {\em to some extent} a burst on the same ion along the $y$ direction. However, this is a tricky matter,
as the phase relation between the two directions might be such that the two bursts would not 
necessarily appear correlated, depending on the spectral resolution and on the burst lifetime itself. 
While  conceiving the appropriate tool to enforce such constraints as rigorously as possible,
we are nonetheless reporting  here some interesting results obtained with the simplest 
sieving procedure outlined above.\\
\indent Direct inspection of Fig.~\ref{f:ILMdef} shows that the number of putative 
ILM excitations increases with temperature. Moreover, it seems that the excitations
that fall on the [110] theoretical curves are much more abundant than the [100] and 
[111] excitations, despite that the theory developed in Ref.~\onlinecite{Nevedrov:2001aa} 
predicted  the [111] modes to be the most stable ones. However, it should be  
remarked that the lifetime $\approx 3 \times 10^{-9}$ s,
predicted in Ref.~\onlinecite{Nevedrov:2001aa} for the [111] modes based on the interaction 
with a (Bose-Einstein) thermal distribution of phonons, exceeds by two orders of magnitude 
the longest lifetimes assigned to a burst in the gap in this study (about 30 ps).\\
\indent The top left panel in Fig.~\ref{f:ILM4panels} compares the site-occupancy 
probabilities relative to the ILM subpopulations to the global site-occupancy 
probability of the whole burst database. The data are well fitted by general chemical
equilibria between two free-energy minima (possibly separated by a number of intermediates),
embodied by expression~\eqref{e:burstEQ}. The corresponding free-energy differences 
are reported in Table~\ref{t:ILM}.
It can be appreciated that putative ILM excitations along [100] and [111]
appear to be rather in the minority with respect 
to {\em generic} burst excitations. Putative [110] modes seem to be more numerous at low 
and intermediate temperature. Nonetheless, the population of these kind of excitations 
seem to increase with temperature as that of the generic bursts, while [100] and [111]
modes appear to be about three orders of magnitude less than generic bursts at 
intermediate temperatures, while surging in number with temperature much more rapidly than 
[110] modes. This is reflected by the best-fit 
value of the enthalpy and entropy differences  (see Table~\ref{t:ILM}). Putative [100] and 
[111] ILM-like bursts seem far easier to excite from the point of view of entropy than 
[110] excitations, explaining the marked temperature dependence of their SOP. It is interesting 
to observe that the predictions made in Ref.~\onlinecite{Sievers:2013aa} for [111] modes
seem to underestimate the excitation entropy difference (4 $k_B$ vs 12 $k_B$), which results
in a reduced temperature dependence of their excitation equilibrium (dashed line in  the
top left panel of Fig.~\ref{f:ILM4panels}). 
%
%
This might indicate that in general at thermal 
equilibrium there might be more excitation channels than merely 
specified by the symmetry-equivalent points at the boundary 
of the Brillouin zone (L points in the case of [111] modes). 
These might reflect interconversion events or mixed-character modes, 
as already suggested in Ref.~\onlinecite{Manley:2011aa}.\\
%
\begin{table}
\caption{\label{t:ILM} Best-fit values of the energy and entropy differences describing 
the equilibrium between energy fluctuations and stabilized bursts according to 
the law~\eqref{e:burstEQ}, with $\Delta f = \Delta \epsilon - T\Delta s$.
The excitations labeled according to different polarizations correspond to the 
sub-populations sieved out at each temperature from the whole ensemble of bursts 
by keeping only the excitations that match the 
corresponding theoretical frequency-amplitude relations taken from 
Ref.~\onlinecite{Nevedrov:2001aa} (see again Fig.~\ref{f:ILMdef}).
}
\begin{ruledtabular}
\begin{tabular}{crr}
%
%
Excitation kind  &    $\Delta\epsilon$  [eV] &   $\Delta s$ $[k_B]$ \\
\hline
All    &  0.54 $\pm$ 0.01   &  9    $\pm$ 0.2  \\
$[100]$  &  1.16 $\pm$ 0.06   &  11.4 $\pm$ 0.8  \\
$[110]$  &  0.32 $\pm$ 0.01   &  2.5  $\pm$ 0.2  \\  
$[111]$  &  1.06 $\pm$ 0.04   &  11.9 $\pm$ 0.5  \\
\end{tabular}
\end{ruledtabular}
\end{table}
%
\indent An analysis of the lifetimes measured for putative ILM-like
excitations also confirms some of the predictions made in Ref.~\onlinecite{Nevedrov:2001aa}
(top right panel in Fig.~\ref{f:ILM4panels}).
Excitations along [100] and [110] display lower-than-average 
lifetimes, while the lifetimes of [111] excitations increase rapidly with 
temperature, to last beyond average bursts at high temperatures.
Interestingly, the lifetimes of [100] and [111] bursts seem to display a marked dependence on
temperature, matched by their rapidly increasing SOP, while [110] excitations
show nearly temperature-independent lifetimes, rhyming with a much more slowly increasing
SOP (top left panel). This seems to point to a less marked nonlinear character
for bursts sieved out along [110]. \\
\indent Amplitudes and energies of bursts seem to trace a consistent picture
(bottom panels in Fig.~\ref{f:ILM4panels}). While along [110], and to a lesser extent along 
[100], the data relative to the putative theoretical sub-populations display trends 
that are consistent with the average behavior of the whole burst database, the [111] 
sub-ensemble demonstrates a  substantially contrasting trend. More specifically, 
excitations selected to lie along the theoretical [111] dispersion law display
systematically higher-than-average energies and larger-than-average 
amplitudes. This is consistent with a more marked nonlinear character of 
these excitations, which in turn upholds the predictions reported in
Ref.~\onlinecite{Nevedrov:2001aa} concerning the markedly higher lifetime of [111] ILMs.

\section{Conclusions and discussion}
\label{sec:conclu}

\noindent In this paper we have introduced a method to resolve transient localization of 
energy in time-frequency space. Our technique is based on continuous wavelet transform of 
velocity time series coupled 
to a threshold-dependent filtering procedure to isolate excitation events from background noise 
in a specific spectral region. A frequency integration in the reference spectral region 
allows us to track the time evolution of the center-of-mass frequency of that region. 
These reduced data, in turn, can be easily exploited to investigate the statistics of 
the burst excitation dynamics. For example,  this procedure can be employed to 
characterize the distribution of the burst lifetimes and investigate the roots of the 
excitation process by looking at the distribution of excitation times (time intervals 
separating consecutive excitation events).\\
\indent As an illustration of our method, we have employed the wavelet-based energy burst 
imaging technique to investigate spontaneous localization of nonlinear modes in the gap of 
NaI crystals at high temperature. Our method allows one to build a database of excitation 
events, and to measure their site-occupancy probability, average lifetime, energy, frequency, 
amplitude and excitation times. It is highly likely that such database contain sub-populations 
corresponding to spontaneous excitation of ILMs, provided a sufficient number of events is 
recorded, {\em i.e.} provided large enough systems are considered and long-enough trajectories 
are simulated.  Overall, the burst database shows rather clearly that the events recorded 
are thermally excited. One way to rationalize the overall excitation equilibrium and kinetics
is in terms of a reaction kinetic scheme involving chemical species equivalents,
representing {\em fluctuations} (F), {\em bursts} (B) along with a variable number of 
intermediates. The numerically measured lifetimes and excitation times suggest that 
such kind of reaction scheme is associated with an energy landscape with as many minima as 
different virtual species. It is possible that this analogy could be pushed even farther 
than this, through the identification of the appropriate collective coordinates (the support 
of the energy landscape), which could allow one to reconstruct the landscape from the 
simulations through standard free-energy calculation algorithms.\\
\indent The problem than one faces in the second logical stage of our method is how to 
single out events corresponding to {\em genuine} ILM excitation, as opposed to generic soft 
nonlinear phonon excitations. We observe that this is a rather formidable task, as the 
fraction of such events is expected to be low, while their polarization and 
localization length can only be guessed from zero-temperature calculations. In this paper,
we have followed a very simple and minimalistic strategy, based explicitly on the zero-temperature
predictions, to sift through the whole burst database at each temperature in the quest for 
ILM events. This procedure seems to succeed, at least partially, in the task of isolating 
events that display a marked nonlinear character. In particular, events selected from the 
burst database by matching the theoretical $T=0$ frequency-amplitude relation for the 
[111] polarization seem to detach the most from the average behavior of the entire databases,
suggesting that at least some of these events might be genuine ILMs along [111]. 
The corresponding site-occupancy probability for these events is described by the 
same theoretical expression as suggested in Ref.~\onlinecite{Sievers:2013aa},
although we find that there might be more excitation pathways for these modes than merely 
specified by the symmetry-equivalent points at the boundary 
of the Brillouin zone (L points). This might reflect interconversion events or 
mixed-character modes, as hinted at in Ref.~\onlinecite{Manley:2011aa}.\\
\indent From a general point of view, it is hard to state whether thermal
populations of ILMs in crystals allow them to be detected and characterized 
directly from equilibrium MD simulations. It is possible that this would require, in general,
some sort of an intrinsically nonlinear pump-probe technique to enhance selectively 
thermal populations of nonlinear excitations. A clever example of amplification and counting 
of ILM excitations is reported in Ref.~\onlinecite{Sato:2005aa}
for quasi-one-dimensional antiferromagnetic lattices,
where an original pump-probe technique based on a four-wave mixing amplification of the 
weak signal from the few large-amplitude ILMs is used to count ILM emission events.
In principle, an ILM generation and steady-state locking techniques such as further
discussed in Ref.~\onlinecite{Wrubel:2005aa} could be implemented numerically 
to produce energy localization in a controlled fashion in atomic lattices at high temperature.\\
\indent In general, ILM localization is expected to be accompanied by a strain 
field (sometimes referred to as the DC component) as a result of odd-order 
anharmonic terms. Moreover, as suggested in Ref.~\onlinecite{Manley:2014aa}, the strain field
associated with thermal excitation of ILMs is expected to take the form of 
planar fault-like structures with an occurrence frequency $f$ of approximately 
one in every ten cells ($f=1/10$).   
However, our method is based on the analysis of velocity time-series.
Therefore, it is insensitive in principle to static distortions associated with the ILM 
displacement fields. Nonetheless, we observe that a {\em spatial} version of our method could be designed
in principle to detect the features of the strain fields associated with ILMs, 
by Gabor transforming spatial-Fourier transformed time series corresponding to specific wavevectors. 
To make contact with the results reported  in Ref.~\onlinecite{Manley:2014aa}, one should 
also consider larger systems including at least twice as many cells in each directions 
than the present study.\\
\indent Although we demonstrated here the power of wavelet-based imaging to investigate the dynamics 
of nonlinear excitations in the gap of NaI crystals, methods of the like can be useful in many
contexts where one wishes to characterize transient energy excitation or energy transfer processes.
The latter kind of phenomena, which is not investigated here, appears to be a promising domain 
of application of our method, both at the classical and quantum level. For example, it would 
be interesting to adopt a tool inspired to our method to characterize the dynamics of 
energy transfer and exciton-phonon interactions in light-harvesting 
complexes.~\cite{Pouyandeh:2017aa,Viani:2014kl,Wu:2010aa} Wavelet-based 
methods could be used to characterize the dynamics of 
vibrational energy transfer~\cite{Yu:2003fk,Leitner:2008gl} 
in many complex system, including biomolecules. For example, coupled to 
pump-probe molecular dynamics approaches~\cite{Sharp:2006aa} 
to investigate in a time-resolved manner 
long-range coupling~\cite{Piazza:2009aa} in frequency space.
These analysis could provide important information as to the 
structural and dynamical determinants of allosteric communication 
in proteins.~\cite{Tsai:2014aa} More generally, our method could make it possible to 
reconstruct the topology of the network of nonlinear interactions 
in a normal-mode space that is dual to the geography of energy 
redistribution in 3D space in many-body systems.~\cite{Allain:2014aa}

%
%

\begin{thebibliography}{87}%
\makeatletter
\providecommand \@ifxundefined [1]{%
 \@ifx{#1\undefined}
}%
\providecommand \@ifnum [1]{%
 \ifnum #1\expandafter \@firstoftwo
 \else \expandafter \@secondoftwo
 \fi
}%
\providecommand \@ifx [1]{%
 \ifx #1\expandafter \@firstoftwo
 \else \expandafter \@secondoftwo
 \fi
}%
\providecommand \natexlab [1]{#1}%
\providecommand \enquote  [1]{``#1''}%
\providecommand \bibnamefont  [1]{#1}%
\providecommand \bibfnamefont [1]{#1}%
\providecommand \citenamefont [1]{#1}%
\providecommand \href@noop [0]{\@secondoftwo}%
\providecommand \href [0]{\begingroup \@sanitize@url \@href}%
\providecommand \@href[1]{\@@startlink{#1}\@@href}%
\providecommand \@@href[1]{\endgroup#1\@@endlink}%
\providecommand \@sanitize@url [0]{\catcode `\\12\catcode `\$12\catcode
  `\&12\catcode `\#12\catcode `\^12\catcode `\_12\catcode `\%12\relax}%
\providecommand \@@startlink[1]{}%
\providecommand \@@endlink[0]{}%
\providecommand \url  [0]{\begingroup\@sanitize@url \@url }%
\providecommand \@url [1]{\endgroup\@href {#1}{\urlprefix }}%
\providecommand \urlprefix  [0]{URL }%
\providecommand \Eprint [0]{\href }%
\providecommand \doibase [0]{http://dx.doi.org/}%
\providecommand \selectlanguage [0]{\@gobble}%
\providecommand \bibinfo  [0]{\@secondoftwo}%
\providecommand \bibfield  [0]{\@secondoftwo}%
\providecommand \translation [1]{[#1]}%
\providecommand \BibitemOpen [0]{}%
\providecommand \bibitemStop [0]{}%
\providecommand \bibitemNoStop [0]{.\EOS\space}%
\providecommand \EOS [0]{\spacefactor3000\relax}%
\providecommand \BibitemShut  [1]{\csname bibitem#1\endcsname}%
\let\auto@bib@innerbib\@empty
\bibitem [{\citenamefont {Mar{\'\i}n}\ \emph {et~al.}(1998)\citenamefont
  {Mar{\'\i}n}, \citenamefont {Aubry},\ and\ \citenamefont
  {Flor{\'\i}a}}]{Marin:1998gj}%
  \BibitemOpen
  \bibfield  {author} {\bibinfo {author} {\bibfnamefont {J.~L.}\ \bibnamefont
  {Mar{\'\i}n}}, \bibinfo {author} {\bibfnamefont {S.}~\bibnamefont {Aubry}}, \
  and\ \bibinfo {author} {\bibfnamefont {L.~M.}\ \bibnamefont {Flor{\'\i}a}},\
  }\bibfield  {booktitle} {\emph {\bibinfo {booktitle} {Proceedings of the
  Conference on Fluctuations, Nonlinearity and Disorder in Condensed Matter and
  Biological Physics}},\ }\href@noop {} {\bibfield  {journal} {\bibinfo
  {journal} {Physica D: Nonlinear Phenomena}\ }\textbf {\bibinfo {volume}
  {113}},\ \bibinfo {pages} {283} (\bibinfo {year} {1998})}\BibitemShut
  {NoStop}%
\bibitem [{\citenamefont {Flach}\ and\ \citenamefont
  {Gorbach}(2008)}]{Flach:2008aa}%
  \BibitemOpen
  \bibfield  {author} {\bibinfo {author} {\bibfnamefont {S.}~\bibnamefont
  {Flach}}\ and\ \bibinfo {author} {\bibfnamefont {A.~V.}\ \bibnamefont
  {Gorbach}},\ }\href@noop {} {\bibfield  {journal} {\bibinfo  {journal}
  {Physics Reports}\ }\textbf {\bibinfo {volume} {467}},\ \bibinfo {pages} {1}
  (\bibinfo {year} {2008})}\BibitemShut {NoStop}%
\bibitem [{\citenamefont {MacKay}\ and\ \citenamefont
  {Aubry}(1994)}]{MacKay:1994aa}%
  \BibitemOpen
  \bibfield  {author} {\bibinfo {author} {\bibfnamefont {R.~S.}\ \bibnamefont
  {MacKay}}\ and\ \bibinfo {author} {\bibfnamefont {S.}~\bibnamefont {Aubry}},\
  }\href@noop {} {\bibfield  {journal} {\bibinfo  {journal} {Nonlinearity}\
  }\textbf {\bibinfo {volume} {7}},\ \bibinfo {pages} {1623} (\bibinfo {year}
  {1994})}\BibitemShut {NoStop}%
\bibitem [{\citenamefont {Sievers}\ and\ \citenamefont
  {Takeno}(1988)}]{Sievers:1988or}%
  \BibitemOpen
  \bibfield  {author} {\bibinfo {author} {\bibfnamefont {A.~J.}\ \bibnamefont
  {Sievers}}\ and\ \bibinfo {author} {\bibfnamefont {S.}~\bibnamefont
  {Takeno}},\ }\href@noop {} {\bibfield  {journal} {\bibinfo  {journal}
  {Physical Review Letters}\ }\textbf {\bibinfo {volume} {61}} (\bibinfo {year}
  {1988})}\BibitemShut {NoStop}%
\bibitem [{\citenamefont {Ford}(1992)}]{Ford:1992aa}%
  \BibitemOpen
  \bibfield  {author} {\bibinfo {author} {\bibfnamefont {J.}~\bibnamefont
  {Ford}},\ }\href@noop {} {\bibfield  {journal} {\bibinfo  {journal} {Physics
  Reports}\ }\textbf {\bibinfo {volume} {213}},\ \bibinfo {pages} {271}
  (\bibinfo {year} {1992})}\BibitemShut {NoStop}%
\bibitem [{\citenamefont {Dmitriev}\ \emph {et~al.}(2015)\citenamefont
  {Dmitriev}, \citenamefont {Chetverikov},\ and\ \citenamefont
  {Velarde}}]{Dmitriev:2015aa}%
  \BibitemOpen
  \bibfield  {author} {\bibinfo {author} {\bibfnamefont {S.~V.}\ \bibnamefont
  {Dmitriev}}, \bibinfo {author} {\bibfnamefont {A.~P.}\ \bibnamefont
  {Chetverikov}}, \ and\ \bibinfo {author} {\bibfnamefont {M.~G.}\ \bibnamefont
  {Velarde}},\ }\href@noop {} {\bibfield  {journal} {\bibinfo  {journal}
  {Physica Status Solidi B}\ }\textbf {\bibinfo {volume} {252}},\ \bibinfo
  {pages} {1682} (\bibinfo {year} {2015})}\BibitemShut {NoStop}%
\bibitem [{\citenamefont {Kiselev}\ and\ \citenamefont
  {Sievers}(1997)}]{Kiselev:1997lr}%
  \BibitemOpen
  \bibfield  {author} {\bibinfo {author} {\bibfnamefont {S.~A.}\ \bibnamefont
  {Kiselev}}\ and\ \bibinfo {author} {\bibfnamefont {A.~J.}\ \bibnamefont
  {Sievers}},\ }\href@noop {} {\bibfield  {journal} {\bibinfo  {journal}
  {Physical Review B}\ }\textbf {\bibinfo {volume} {55}} (\bibinfo {year}
  {1997})}\BibitemShut {NoStop}%
\bibitem [{\citenamefont {Hizhnyakov}\ \emph
  {et~al.}(2016{\natexlab{a}})\citenamefont {Hizhnyakov}, \citenamefont {Haas},
  \citenamefont {Klopov},\ and\ \citenamefont {Shelkan}}]{Hizhnyakov:2016ab}%
  \BibitemOpen
  \bibfield  {author} {\bibinfo {author} {\bibfnamefont {V.}~\bibnamefont
  {Hizhnyakov}}, \bibinfo {author} {\bibfnamefont {M.}~\bibnamefont {Haas}},
  \bibinfo {author} {\bibfnamefont {M.}~\bibnamefont {Klopov}}, \ and\ \bibinfo
  {author} {\bibfnamefont {A.}~\bibnamefont {Shelkan}},\ }\href@noop {}
  {\bibfield  {journal} {\bibinfo  {journal} {Letters on Materials}\ }\textbf
  {\bibinfo {volume} {6}},\ \bibinfo {pages} {61} (\bibinfo {year}
  {2016}{\natexlab{a}})}\BibitemShut {NoStop}%
\bibitem [{\citenamefont {Marin}\ \emph {et~al.}(2001)\citenamefont {Marin},
  \citenamefont {Russell},\ and\ \citenamefont {Eilbeck}}]{Marin:2001fj}%
  \BibitemOpen
  \bibfield  {author} {\bibinfo {author} {\bibfnamefont {J.~L.}\ \bibnamefont
  {Marin}}, \bibinfo {author} {\bibfnamefont {F.~M.}\ \bibnamefont {Russell}},
  \ and\ \bibinfo {author} {\bibfnamefont {J.~C.}\ \bibnamefont {Eilbeck}},\
  }\href
  {http://www.sciencedirect.com/science/article/B6TVM-42HFSG5-14/2/c3170efd535c33ed6e1d4043d295d4cf}
  {\bibfield  {journal} {\bibinfo  {journal} {Physics Letters A}\ }\textbf
  {\bibinfo {volume} {281}},\ \bibinfo {pages} {21} (\bibinfo {year}
  {2001})}\BibitemShut {NoStop}%
\bibitem [{\citenamefont {Barani}\ \emph {et~al.}(2017)\citenamefont {Barani},
  \citenamefont {Korznikova}, \citenamefont {Chetverikov}, \citenamefont
  {Zhou},\ and\ \citenamefont {Dmitriev}}]{Barani:2017aa}%
  \BibitemOpen
  \bibfield  {author} {\bibinfo {author} {\bibfnamefont {E.}~\bibnamefont
  {Barani}}, \bibinfo {author} {\bibfnamefont {E.~A.}\ \bibnamefont
  {Korznikova}}, \bibinfo {author} {\bibfnamefont {A.~P.}\ \bibnamefont
  {Chetverikov}}, \bibinfo {author} {\bibfnamefont {K.}~\bibnamefont {Zhou}}, \
  and\ \bibinfo {author} {\bibfnamefont {S.~V.}\ \bibnamefont {Dmitriev}},\
  }\href@noop {} {\bibfield  {journal} {\bibinfo  {journal} {Physics Letters
  A}\ }\textbf {\bibinfo {volume} {381}},\ \bibinfo {pages} {3553} (\bibinfo
  {year} {2017})}\BibitemShut {NoStop}%
\bibitem [{\citenamefont {Evazzade}\ \emph {et~al.}(2018)\citenamefont
  {Evazzade}, \citenamefont {Roknabadi}, \citenamefont {Behdani}, \citenamefont
  {Moosavi}, \citenamefont {Xiong}, \citenamefont {Zhou},\ and\ \citenamefont
  {Dmitriev}}]{Evazzade:2018aa}%
  \BibitemOpen
  \bibfield  {author} {\bibinfo {author} {\bibfnamefont {I.}~\bibnamefont
  {Evazzade}}, \bibinfo {author} {\bibfnamefont {M.~R.}\ \bibnamefont
  {Roknabadi}}, \bibinfo {author} {\bibfnamefont {M.}~\bibnamefont {Behdani}},
  \bibinfo {author} {\bibfnamefont {F.}~\bibnamefont {Moosavi}}, \bibinfo
  {author} {\bibfnamefont {D.}~\bibnamefont {Xiong}}, \bibinfo {author}
  {\bibfnamefont {K.}~\bibnamefont {Zhou}}, \ and\ \bibinfo {author}
  {\bibfnamefont {S.~V.}\ \bibnamefont {Dmitriev}},\ }\href@noop {} {\bibfield
  {journal} {\bibinfo  {journal} {European Physical Journal B}\ }\textbf
  {\bibinfo {volume} {91}},\ \bibinfo {pages} {163} (\bibinfo {year}
  {2018})}\BibitemShut {NoStop}%
\bibitem [{\citenamefont {Fraile}\ \emph {et~al.}(2016)\citenamefont {Fraile},
  \citenamefont {Koukaras}, \citenamefont {Papagelis}, \citenamefont
  {Lazarides},\ and\ \citenamefont {Tsironis}}]{Fraile:2016aa}%
  \BibitemOpen
  \bibfield  {author} {\bibinfo {author} {\bibfnamefont {A.}~\bibnamefont
  {Fraile}}, \bibinfo {author} {\bibfnamefont {E.~N.}\ \bibnamefont
  {Koukaras}}, \bibinfo {author} {\bibfnamefont {K.}~\bibnamefont {Papagelis}},
  \bibinfo {author} {\bibfnamefont {N.}~\bibnamefont {Lazarides}}, \ and\
  \bibinfo {author} {\bibfnamefont {G.~P.}\ \bibnamefont {Tsironis}},\
  }\href@noop {} {\bibfield  {journal} {\bibinfo  {journal} {Chaos Solitons \&
  Fractals}\ }\textbf {\bibinfo {volume} {87}},\ \bibinfo {pages} {262}
  (\bibinfo {year} {2016})}\BibitemShut {NoStop}%
\bibitem [{\citenamefont {Hizhnyakov}\ \emph
  {et~al.}(2016{\natexlab{b}})\citenamefont {Hizhnyakov}, \citenamefont
  {Klopov},\ and\ \citenamefont {Shelkan}}]{Hizhnyakov:2016aa}%
  \BibitemOpen
  \bibfield  {author} {\bibinfo {author} {\bibfnamefont {V.}~\bibnamefont
  {Hizhnyakov}}, \bibinfo {author} {\bibfnamefont {M.}~\bibnamefont {Klopov}},
  \ and\ \bibinfo {author} {\bibfnamefont {A.}~\bibnamefont {Shelkan}},\
  }\href@noop {} {\bibfield  {journal} {\bibinfo  {journal} {Physics Letters
  A}\ }\textbf {\bibinfo {volume} {380}},\ \bibinfo {pages} {1075} (\bibinfo
  {year} {2016}{\natexlab{b}})}\BibitemShut {NoStop}%
\bibitem [{\citenamefont {Murzaev}\ \emph {et~al.}(2017)\citenamefont
  {Murzaev}, \citenamefont {Bachurin}, \citenamefont {Korznikova},\ and\
  \citenamefont {Dmitriev}}]{Murzaev:2017aa}%
  \BibitemOpen
  \bibfield  {author} {\bibinfo {author} {\bibfnamefont {R.~T.}\ \bibnamefont
  {Murzaev}}, \bibinfo {author} {\bibfnamefont {D.~V.}\ \bibnamefont
  {Bachurin}}, \bibinfo {author} {\bibfnamefont {E.~A.}\ \bibnamefont
  {Korznikova}}, \ and\ \bibinfo {author} {\bibfnamefont {S.~V.}\ \bibnamefont
  {Dmitriev}},\ }\href@noop {} {\bibfield  {journal} {\bibinfo  {journal}
  {Physics Letters A}\ }\textbf {\bibinfo {volume} {381}},\ \bibinfo {pages}
  {1003} (\bibinfo {year} {2017})}\BibitemShut {NoStop}%
\bibitem [{\citenamefont {Sepulchre}\ and\ \citenamefont
  {MacKay}(1998)}]{Sepulchre:1998sp}%
  \BibitemOpen
  \bibfield  {author} {\bibinfo {author} {\bibfnamefont {J.~A.}\ \bibnamefont
  {Sepulchre}}\ and\ \bibinfo {author} {\bibfnamefont {R.~S.}\ \bibnamefont
  {MacKay}},\ }\href@noop {} {\bibfield  {journal} {\bibinfo  {journal}
  {Physica D: Nonlinear Phenomena}\ }\textbf {\bibinfo {volume} {113}},\
  \bibinfo {pages} {342} (\bibinfo {year} {1998})}\BibitemShut {NoStop}%
\bibitem [{\citenamefont {Kopidakis}\ and\ \citenamefont
  {Aubry}(1999)}]{Kopidakis:1999fk}%
  \BibitemOpen
  \bibfield  {author} {\bibinfo {author} {\bibfnamefont {G.}~\bibnamefont
  {Kopidakis}}\ and\ \bibinfo {author} {\bibfnamefont {S.}~\bibnamefont
  {Aubry}},\ }\href@noop {} {\bibfield  {journal} {\bibinfo  {journal} {Physica
  D: Nonlinear Phenomena}\ }\textbf {\bibinfo {volume} {130}},\ \bibinfo
  {pages} {155} (\bibinfo {year} {1999})}\BibitemShut {NoStop}%
\bibitem [{\citenamefont {Kopidakis}\ and\ \citenamefont
  {Aubry}(2000)}]{Kopidakis:2000lr}%
  \BibitemOpen
  \bibfield  {author} {\bibinfo {author} {\bibfnamefont {G.}~\bibnamefont
  {Kopidakis}}\ and\ \bibinfo {author} {\bibfnamefont {S.}~\bibnamefont
  {Aubry}},\ }\href@noop {} {\bibfield  {journal} {\bibinfo  {journal} {Physica
  D: Nonlinear Phenomena}\ }\textbf {\bibinfo {volume} {139}},\ \bibinfo
  {pages} {247} (\bibinfo {year} {2000})}\BibitemShut {NoStop}%
\bibitem [{\citenamefont {Peyrard}(1995)}]{breath-macromol}%
  \BibitemOpen
  \bibinfo {editor} {\bibfnamefont {M.}~\bibnamefont {Peyrard}},\ ed.,\
  \href@noop {} {\emph {\bibinfo {title} {Nonlinear excitations in
  biomolecules}}}\ (\bibinfo  {publisher} {Springer: Berlin, Heidelberg},\
  \bibinfo {year} {1995})\BibitemShut {NoStop}%
\bibitem [{\citenamefont {Juanico}\ \emph {et~al.}(2007)\citenamefont
  {Juanico}, \citenamefont {Sanejouand}, \citenamefont {Piazza},\ and\
  \citenamefont {De~Los~Rios}}]{Juanico:2007ng}%
  \BibitemOpen
  \bibfield  {author} {\bibinfo {author} {\bibfnamefont {B.}~\bibnamefont
  {Juanico}}, \bibinfo {author} {\bibfnamefont {Y.~H.}\ \bibnamefont
  {Sanejouand}}, \bibinfo {author} {\bibfnamefont {F.}~\bibnamefont {Piazza}},
  \ and\ \bibinfo {author} {\bibfnamefont {P.}~\bibnamefont {De~Los~Rios}},\
  }\href@noop {} {\bibfield  {journal} {\bibinfo  {journal} {Physical Review
  Letters}\ }\textbf {\bibinfo {volume} {99}},\ \bibinfo {pages} {238104}
  (\bibinfo {year} {2007})}\BibitemShut {NoStop}%
\bibitem [{\citenamefont {Piazza}\ and\ \citenamefont
  {Sanejouand}(2008)}]{Piazza:2008mb}%
  \BibitemOpen
  \bibfield  {author} {\bibinfo {author} {\bibfnamefont {F.}~\bibnamefont
  {Piazza}}\ and\ \bibinfo {author} {\bibfnamefont {Y.-H.}\ \bibnamefont
  {Sanejouand}},\ }\href@noop {} {\bibfield  {journal} {\bibinfo  {journal}
  {Physical Biology}\ }\textbf {\bibinfo {volume} {5}},\ \bibinfo {pages}
  {026001} (\bibinfo {year} {2008})}\BibitemShut {NoStop}%
\bibitem [{\citenamefont {Iubini}\ \emph {et~al.}(2013)\citenamefont {Iubini},
  \citenamefont {Franzosi}, \citenamefont {Livi}, \citenamefont {Oppo},\ and\
  \citenamefont {Politi}}]{Iubini:2013aa}%
  \BibitemOpen
  \bibfield  {author} {\bibinfo {author} {\bibfnamefont {S.}~\bibnamefont
  {Iubini}}, \bibinfo {author} {\bibfnamefont {R.}~\bibnamefont {Franzosi}},
  \bibinfo {author} {\bibfnamefont {R.}~\bibnamefont {Livi}}, \bibinfo {author}
  {\bibfnamefont {G.-L.}\ \bibnamefont {Oppo}}, \ and\ \bibinfo {author}
  {\bibfnamefont {A.}~\bibnamefont {Politi}},\ }\href@noop {} {\bibfield
  {journal} {\bibinfo  {journal} {New Journal of Physics}\ }\textbf {\bibinfo
  {volume} {15}},\ \bibinfo {pages} {023032} (\bibinfo {year}
  {2013})}\BibitemShut {NoStop}%
\bibitem [{\citenamefont {Dubinko}\ \emph {et~al.}(2015)\citenamefont
  {Dubinko}, \citenamefont {Archilla}, \citenamefont {Dmitriev},\ and\
  \citenamefont {Hizhnyakov}}]{Dubinko:2015aa}%
  \BibitemOpen
  \bibfield  {author} {\bibinfo {author} {\bibfnamefont {V.~I.}\ \bibnamefont
  {Dubinko}}, \bibinfo {author} {\bibfnamefont {J.~F.~R.}\ \bibnamefont
  {Archilla}}, \bibinfo {author} {\bibfnamefont {S.~V.}\ \bibnamefont
  {Dmitriev}}, \ and\ \bibinfo {author} {\bibfnamefont {V.}~\bibnamefont
  {Hizhnyakov}},\ }in\ \href@noop {} {\emph {\bibinfo {booktitle} {Quodons in
  mica: nonlinear localized travelling excitations in crystals}}},\ \bibinfo
  {series} {Springer Series in Materials Science}, Vol.\ \bibinfo {volume}
  {221},\ \bibinfo {editor} {edited by\ \bibinfo {editor} {\bibfnamefont
  {J.}~\bibnamefont {Archilla}}, \bibinfo {editor} {\bibfnamefont
  {N.}~\bibnamefont {Jimenez}}, \bibinfo {editor} {\bibfnamefont
  {V.}~\bibnamefont {SanchezMorcillo}}, \ and\ \bibinfo {editor} {\bibfnamefont
  {L.}~\bibnamefont {GarciaRaff}}}\ (\bibinfo  {publisher} {Springer},\
  \bibinfo {year} {2015})\ pp.\ \bibinfo {pages} {381--398}\BibitemShut
  {NoStop}%
\bibitem [{\citenamefont {Dubinko}\ and\ \citenamefont
  {Laptev}(2016)}]{Dubinko:2016aa}%
  \BibitemOpen
  \bibfield  {author} {\bibinfo {author} {\bibfnamefont {V.~I.}\ \bibnamefont
  {Dubinko}}\ and\ \bibinfo {author} {\bibfnamefont {D.~V.}\ \bibnamefont
  {Laptev}},\ }\href@noop {} {\bibfield  {journal} {\bibinfo  {journal}
  {Letters on Materials}\ }\textbf {\bibinfo {volume} {6}},\ \bibinfo {pages}
  {16} (\bibinfo {year} {2016})}\BibitemShut {NoStop}%
\bibitem [{\citenamefont {Dubinko}\ and\ \citenamefont
  {Piazza}(2014)}]{Dubinko:2014aa}%
  \BibitemOpen
  \bibfield  {author} {\bibinfo {author} {\bibfnamefont {V.~I.}\ \bibnamefont
  {Dubinko}}\ and\ \bibinfo {author} {\bibfnamefont {F.}~\bibnamefont
  {Piazza}},\ }\href@noop {} {\bibfield  {journal} {\bibinfo  {journal}
  {Letters on Materials}\ }\textbf {\bibinfo {volume} {4}},\ \bibinfo {pages}
  {273} (\bibinfo {year} {2014})}\BibitemShut {NoStop}%
\bibitem [{\citenamefont {Ivanchenko}\ \emph {et~al.}(2004)\citenamefont
  {Ivanchenko}, \citenamefont {Kanakov}, \citenamefont {Shalfeev},\ and\
  \citenamefont {Flach}}]{Ivanchenko:2004aa}%
  \BibitemOpen
  \bibfield  {author} {\bibinfo {author} {\bibfnamefont {M.~V.}\ \bibnamefont
  {Ivanchenko}}, \bibinfo {author} {\bibfnamefont {O.~I.}\ \bibnamefont
  {Kanakov}}, \bibinfo {author} {\bibfnamefont {V.~D.}\ \bibnamefont
  {Shalfeev}}, \ and\ \bibinfo {author} {\bibfnamefont {S.}~\bibnamefont
  {Flach}},\ }\href@noop {} {\bibfield  {journal} {\bibinfo  {journal} {Physica
  D: Nonlinear Phenomena}\ }\textbf {\bibinfo {volume} {198}},\ \bibinfo
  {pages} {120} (\bibinfo {year} {2004})}\BibitemShut {NoStop}%
\bibitem [{\citenamefont {Eleftheriou}\ and\ \citenamefont
  {Flach}(2005)}]{Eleftheriou:2005aa}%
  \BibitemOpen
  \bibfield  {author} {\bibinfo {author} {\bibfnamefont {M.}~\bibnamefont
  {Eleftheriou}}\ and\ \bibinfo {author} {\bibfnamefont {S.}~\bibnamefont
  {Flach}},\ }\href@noop {} {\bibfield  {journal} {\bibinfo  {journal} {Physica
  D: Nonlinear Phenomena}\ }\textbf {\bibinfo {volume} {202}},\ \bibinfo
  {pages} {142} (\bibinfo {year} {2005})}\BibitemShut {NoStop}%
\bibitem [{\citenamefont {Ming}\ \emph {et~al.}(2017)\citenamefont {Ming},
  \citenamefont {Ling}, \citenamefont {Li},\ and\ \citenamefont
  {Ding}}]{Ming:2017aa}%
  \BibitemOpen
  \bibfield  {author} {\bibinfo {author} {\bibfnamefont {Y.}~\bibnamefont
  {Ming}}, \bibinfo {author} {\bibfnamefont {D.~B.}\ \bibnamefont {Ling}},
  \bibinfo {author} {\bibfnamefont {H.~M.}\ \bibnamefont {Li}}, \ and\ \bibinfo
  {author} {\bibfnamefont {Z.~J.}\ \bibnamefont {Ding}},\ }\href@noop {}
  {\bibfield  {journal} {\bibinfo  {journal} {Chaos}\ }\textbf {\bibinfo
  {volume} {27}},\ \bibinfo {pages} {Article number 063106} (\bibinfo {year}
  {2017})}\BibitemShut {NoStop}%
\bibitem [{\citenamefont {Baimova}\ \emph {et~al.}(2017)\citenamefont
  {Baimova}, \citenamefont {Murzaev},\ and\ \citenamefont
  {Rudskoy}}]{Baimova:2017aa}%
  \BibitemOpen
  \bibfield  {author} {\bibinfo {author} {\bibfnamefont {J.~A.}\ \bibnamefont
  {Baimova}}, \bibinfo {author} {\bibfnamefont {R.~T.}\ \bibnamefont
  {Murzaev}}, \ and\ \bibinfo {author} {\bibfnamefont {A.~I.}\ \bibnamefont
  {Rudskoy}},\ }\href@noop {} {\bibfield  {journal} {\bibinfo  {journal}
  {Physics Letters A}\ }\textbf {\bibinfo {volume} {381}},\ \bibinfo {pages}
  {3049} (\bibinfo {year} {2017})}\BibitemShut {NoStop}%
\bibitem [{\citenamefont {Baimova}\ \emph {et~al.}(2016)\citenamefont
  {Baimova}, \citenamefont {Murzaev}, \citenamefont {Lobzenko}, \citenamefont
  {Dmitriev},\ and\ \citenamefont {Zhou}}]{Baimova:2016aa}%
  \BibitemOpen
  \bibfield  {author} {\bibinfo {author} {\bibfnamefont {J.~A.}\ \bibnamefont
  {Baimova}}, \bibinfo {author} {\bibfnamefont {R.~T.}\ \bibnamefont
  {Murzaev}}, \bibinfo {author} {\bibfnamefont {I.~P.}\ \bibnamefont
  {Lobzenko}}, \bibinfo {author} {\bibfnamefont {S.~V.}\ \bibnamefont
  {Dmitriev}}, \ and\ \bibinfo {author} {\bibfnamefont {K.}~\bibnamefont
  {Zhou}},\ }\href@noop {} {\bibfield  {journal} {\bibinfo  {journal} {Journal
  of Experimental and Theoretical Physics}\ }\textbf {\bibinfo {volume}
  {122}},\ \bibinfo {pages} {869} (\bibinfo {year} {2016})}\BibitemShut
  {NoStop}%
\bibitem [{\citenamefont {Kistanov}\ and\ \citenamefont
  {Dmitriev}(2012)}]{Kistanov:2012aa}%
  \BibitemOpen
  \bibfield  {author} {\bibinfo {author} {\bibfnamefont {a.~a.}\ \bibnamefont
  {Kistanov}}\ and\ \bibinfo {author} {\bibfnamefont {S.~V.}\ \bibnamefont
  {Dmitriev}},\ }\href {\doibase 10.1134/S1063783412080148} {\bibfield
  {journal} {\bibinfo  {journal} {Physics of the Solid State}\ }\textbf
  {\bibinfo {volume} {54}},\ \bibinfo {pages} {1648} (\bibinfo {year}
  {2012})}\BibitemShut {NoStop}%
\bibitem [{\citenamefont {Binder}\ \emph {et~al.}(2000)\citenamefont {Binder},
  \citenamefont {Abraimov}, \citenamefont {Ustinov}, \citenamefont {Flach},\
  and\ \citenamefont {Zolotaryuk}}]{Binder:2000ih}%
  \BibitemOpen
  \bibfield  {author} {\bibinfo {author} {\bibfnamefont {P.}~\bibnamefont
  {Binder}}, \bibinfo {author} {\bibfnamefont {D.}~\bibnamefont {Abraimov}},
  \bibinfo {author} {\bibfnamefont {A.~V.}\ \bibnamefont {Ustinov}}, \bibinfo
  {author} {\bibfnamefont {S.}~\bibnamefont {Flach}}, \ and\ \bibinfo {author}
  {\bibfnamefont {Y.}~\bibnamefont {Zolotaryuk}},\ }\href@noop {} {\bibfield
  {journal} {\bibinfo  {journal} {Physical Review Letters}\ }\textbf {\bibinfo
  {volume} {84}} (\bibinfo {year} {2000})}\BibitemShut {NoStop}%
\bibitem [{\citenamefont {Trias}\ \emph {et~al.}(2000)\citenamefont {Trias},
  \citenamefont {Mazo},\ and\ \citenamefont {Orlando}}]{Trias:2000ej}%
  \BibitemOpen
  \bibfield  {author} {\bibinfo {author} {\bibfnamefont {E.}~\bibnamefont
  {Trias}}, \bibinfo {author} {\bibfnamefont {J.~J.}\ \bibnamefont {Mazo}}, \
  and\ \bibinfo {author} {\bibfnamefont {T.~P.}\ \bibnamefont {Orlando}},\
  }\href@noop {} {\bibfield  {journal} {\bibinfo  {journal} {Physical Review
  Letters}\ }\textbf {\bibinfo {volume} {84}} (\bibinfo {year}
  {2000})}\BibitemShut {NoStop}%
\bibitem [{\citenamefont {Steeds}\ \emph {et~al.}(1993)\citenamefont {Steeds},
  \citenamefont {Russell},\ and\ \citenamefont {Vine}}]{Steeds:1993aa}%
  \BibitemOpen
  \bibfield  {author} {\bibinfo {author} {\bibfnamefont {J.~W.}\ \bibnamefont
  {Steeds}}, \bibinfo {author} {\bibfnamefont {F.~M.}\ \bibnamefont {Russell}},
  \ and\ \bibinfo {author} {\bibfnamefont {W.~J.}\ \bibnamefont {Vine}},\
  }\href@noop {} {\bibfield  {journal} {\bibinfo  {journal} {Optik}\ }\textbf
  {\bibinfo {volume} {92}},\ \bibinfo {pages} {149} (\bibinfo {year}
  {1993})}\BibitemShut {NoStop}%
\bibitem [{\citenamefont {Russell}(2015)}]{Russell:2015aa}%
  \BibitemOpen
  \bibfield  {author} {\bibinfo {author} {\bibfnamefont {F.~M.}\ \bibnamefont
  {Russell}},\ }in\ \href@noop {} {\emph {\bibinfo {booktitle} {Quodons in
  mica: nonlinear localized travelling excitations in crystals}}},\ \bibinfo
  {series} {{Springer Series in Materials Science}}, Vol.\ \bibinfo {volume}
  {221},\ \bibinfo {editor} {edited by\ \bibinfo {editor} {\bibfnamefont
  {J.}~\bibnamefont {Archilla}}, \bibinfo {editor} {\bibfnamefont
  {N.}~\bibnamefont {Jimenez}}, \bibinfo {editor} {\bibfnamefont
  {V.}~\bibnamefont {Sanchez~Morcillo}}, \ and\ \bibinfo {editor}
  {\bibfnamefont {L.}~\bibnamefont {Garcia~Raff}}}\ (\bibinfo  {publisher}
  {Springer: Cham},\ \bibinfo {address} {Cham},\ \bibinfo {year} {2015})\ pp.\
  \bibinfo {pages} {3--33}\BibitemShut {NoStop}%
\bibitem [{\citenamefont {Marin}\ \emph {et~al.}(1998)\citenamefont {Marin},
  \citenamefont {Eilbeck},\ and\ \citenamefont {Russell}}]{Marin:1998aa}%
  \BibitemOpen
  \bibfield  {author} {\bibinfo {author} {\bibfnamefont {J.~L.}\ \bibnamefont
  {Marin}}, \bibinfo {author} {\bibfnamefont {J.~C.}\ \bibnamefont {Eilbeck}},
  \ and\ \bibinfo {author} {\bibfnamefont {F.~M.}\ \bibnamefont {Russell}},\
  }\href@noop {} {\bibfield  {journal} {\bibinfo  {journal} {Physics Letters,
  Section A: General, Atomic and Solid State Physics}\ }\textbf {\bibinfo
  {volume} {248}},\ \bibinfo {pages} {225} (\bibinfo {year}
  {1998})}\BibitemShut {NoStop}%
\bibitem [{\citenamefont {Russell}\ \emph {et~al.}(2017)\citenamefont
  {Russell}, \citenamefont {Archilla}, \citenamefont {Frutos},\ and\
  \citenamefont {Medina-Carrasco}}]{Russell:2017aa}%
  \BibitemOpen
  \bibfield  {author} {\bibinfo {author} {\bibfnamefont {F.~M.}\ \bibnamefont
  {Russell}}, \bibinfo {author} {\bibfnamefont {J.~F.}\ \bibnamefont
  {Archilla}}, \bibinfo {author} {\bibfnamefont {F.}~\bibnamefont {Frutos}}, \
  and\ \bibinfo {author} {\bibfnamefont {S.}~\bibnamefont {Medina-Carrasco}},\
  }\href@noop {} {\bibfield  {journal} {\bibinfo  {journal} {EPL}\ }\textbf
  {\bibinfo {volume} {120}},\ \bibinfo {pages} {46001} (\bibinfo {year}
  {2017})}\BibitemShut {NoStop}%
\bibitem [{\citenamefont {Manley}\ \emph {et~al.}(2006)\citenamefont {Manley},
  \citenamefont {Yethiraj}, \citenamefont {Sinn}, \citenamefont {Volz},
  \citenamefont {Alatas}, \citenamefont {Lashley}, \citenamefont {Hults},
  \citenamefont {Lander},\ and\ \citenamefont {Smith}}]{Manley:2006aa}%
  \BibitemOpen
  \bibfield  {author} {\bibinfo {author} {\bibfnamefont {M.~E.}\ \bibnamefont
  {Manley}}, \bibinfo {author} {\bibfnamefont {M.}~\bibnamefont {Yethiraj}},
  \bibinfo {author} {\bibfnamefont {H.}~\bibnamefont {Sinn}}, \bibinfo {author}
  {\bibfnamefont {H.~M.}\ \bibnamefont {Volz}}, \bibinfo {author}
  {\bibfnamefont {A.}~\bibnamefont {Alatas}}, \bibinfo {author} {\bibfnamefont
  {J.~C.}\ \bibnamefont {Lashley}}, \bibinfo {author} {\bibfnamefont {W.~L.}\
  \bibnamefont {Hults}}, \bibinfo {author} {\bibfnamefont {G.~H.}\ \bibnamefont
  {Lander}}, \ and\ \bibinfo {author} {\bibfnamefont {J.~L.}\ \bibnamefont
  {Smith}},\ }\href@noop {} {\bibfield  {journal} {\bibinfo  {journal} {Phys.
  Rev. Lett.}\ }\textbf {\bibinfo {volume} {96}},\ \bibinfo {pages} {125501}
  (\bibinfo {year} {2006})}\BibitemShut {NoStop}%
\bibitem [{\citenamefont {Manley}\ \emph {et~al.}(2008)\citenamefont {Manley},
  \citenamefont {Alatas}, \citenamefont {Trouw}, \citenamefont {Leu},
  \citenamefont {Lynn}, \citenamefont {Chen},\ and\ \citenamefont
  {Hults}}]{Manley:2008aa}%
  \BibitemOpen
  \bibfield  {author} {\bibinfo {author} {\bibfnamefont {M.~E.}\ \bibnamefont
  {Manley}}, \bibinfo {author} {\bibfnamefont {A.}~\bibnamefont {Alatas}},
  \bibinfo {author} {\bibfnamefont {F.}~\bibnamefont {Trouw}}, \bibinfo
  {author} {\bibfnamefont {B.~M.}\ \bibnamefont {Leu}}, \bibinfo {author}
  {\bibfnamefont {J.~W.}\ \bibnamefont {Lynn}}, \bibinfo {author}
  {\bibfnamefont {Y.}~\bibnamefont {Chen}}, \ and\ \bibinfo {author}
  {\bibfnamefont {W.~L.}\ \bibnamefont {Hults}},\ }\href@noop {} {\bibfield
  {journal} {\bibinfo  {journal} {Phys. Rev. B}\ }\textbf {\bibinfo {volume}
  {77}},\ \bibinfo {pages} {214305} (\bibinfo {year} {2008})}\BibitemShut
  {NoStop}%
\bibitem [{\citenamefont {Murzaev}\ \emph {et~al.}(2016)\citenamefont
  {Murzaev}, \citenamefont {Babicheva}, \citenamefont {Zhou}, \citenamefont
  {Korznikova}, \citenamefont {Fomin}, \citenamefont {Dubinko},\ and\
  \citenamefont {Dmitriev}}]{Murzaev:2016aa}%
  \BibitemOpen
  \bibfield  {author} {\bibinfo {author} {\bibfnamefont {R.~T.}\ \bibnamefont
  {Murzaev}}, \bibinfo {author} {\bibfnamefont {R.~I.}\ \bibnamefont
  {Babicheva}}, \bibinfo {author} {\bibfnamefont {K.}~\bibnamefont {Zhou}},
  \bibinfo {author} {\bibfnamefont {E.~A.}\ \bibnamefont {Korznikova}},
  \bibinfo {author} {\bibfnamefont {S.~Y.}\ \bibnamefont {Fomin}}, \bibinfo
  {author} {\bibfnamefont {V.~I.}\ \bibnamefont {Dubinko}}, \ and\ \bibinfo
  {author} {\bibfnamefont {S.~V.}\ \bibnamefont {Dmitriev}},\ }\href@noop {}
  {\bibfield  {journal} {\bibinfo  {journal} {European Physical Journal B}\
  }\textbf {\bibinfo {volume} {89}},\ \bibinfo {pages} {168} (\bibinfo {year}
  {2016})}\BibitemShut {NoStop}%
\bibitem [{\citenamefont {Manley}\ \emph {et~al.}(2009)\citenamefont {Manley},
  \citenamefont {Sievers}, \citenamefont {Lynn}, \citenamefont {Kiselev},
  \citenamefont {Agladze}, \citenamefont {Chen}, \citenamefont {Llobet},\ and\
  \citenamefont {Alatas}}]{Manley:2009aa}%
  \BibitemOpen
  \bibfield  {author} {\bibinfo {author} {\bibfnamefont {M.~E.}\ \bibnamefont
  {Manley}}, \bibinfo {author} {\bibfnamefont {A.~J.}\ \bibnamefont {Sievers}},
  \bibinfo {author} {\bibfnamefont {J.~W.}\ \bibnamefont {Lynn}}, \bibinfo
  {author} {\bibfnamefont {S.~A.}\ \bibnamefont {Kiselev}}, \bibinfo {author}
  {\bibfnamefont {N.~I.}\ \bibnamefont {Agladze}}, \bibinfo {author}
  {\bibfnamefont {Y.}~\bibnamefont {Chen}}, \bibinfo {author} {\bibfnamefont
  {A.}~\bibnamefont {Llobet}}, \ and\ \bibinfo {author} {\bibfnamefont
  {A.}~\bibnamefont {Alatas}},\ }\href@noop {} {\bibfield  {journal} {\bibinfo
  {journal} {Physical Review B}\ }\textbf {\bibinfo {volume} {79}},\ \bibinfo
  {pages} {134304} (\bibinfo {year} {2009})}\BibitemShut {NoStop}%
\bibitem [{\citenamefont {Sievers}\ \emph {et~al.}(2013)\citenamefont
  {Sievers}, \citenamefont {Sato}, \citenamefont {Page},\ and\ \citenamefont
  {R{\"{o}}ssler}}]{Sievers:2013aa}%
  \BibitemOpen
  \bibfield  {author} {\bibinfo {author} {\bibfnamefont {A.~J.}\ \bibnamefont
  {Sievers}}, \bibinfo {author} {\bibfnamefont {M.}~\bibnamefont {Sato}},
  \bibinfo {author} {\bibfnamefont {J.~B.}\ \bibnamefont {Page}}, \ and\
  \bibinfo {author} {\bibfnamefont {T.}~\bibnamefont {R{\"{o}}ssler}},\
  }\href@noop {} {\bibfield  {journal} {\bibinfo  {journal} {Physical Review B
  - Condensed Matter and Materials Physics}\ }\textbf {\bibinfo {volume} {88}}
  (\bibinfo {year} {2013})}\BibitemShut {NoStop}%
\bibitem [{\citenamefont {Khadeeva}\ and\ \citenamefont
  {Dmitriev}(2010)}]{Khadeeva:2010aa}%
  \BibitemOpen
  \bibfield  {author} {\bibinfo {author} {\bibfnamefont {L.~Z.}\ \bibnamefont
  {Khadeeva}}\ and\ \bibinfo {author} {\bibfnamefont {S.~V.}\ \bibnamefont
  {Dmitriev}},\ }\href {\doibase 10.1103/PhysRevB.81.214306} {\bibfield
  {journal} {\bibinfo  {journal} {Physical Review B}\ }\textbf {\bibinfo
  {volume} {81}},\ \bibinfo {pages} {214306} (\bibinfo {year}
  {2010})}\BibitemShut {NoStop}%
\bibitem [{\citenamefont {Nevedrov}\ \emph {et~al.}(2001)\citenamefont
  {Nevedrov}, \citenamefont {Hizhnyakov},\ and\ \citenamefont
  {Sievers}}]{Nevedrov:2001aa}%
  \BibitemOpen
  \bibfield  {author} {\bibinfo {author} {\bibfnamefont {D.}~\bibnamefont
  {Nevedrov}}, \bibinfo {author} {\bibfnamefont {V.}~\bibnamefont
  {Hizhnyakov}}, \ and\ \bibinfo {author} {\bibfnamefont {A.~J.}\ \bibnamefont
  {Sievers}},\ }\enquote {\bibinfo {title} {Anharmonic gap modes in alkali
  halides},}\ in\ \href@noop {} {\emph {\bibinfo {booktitle} {Vibronic
  Interactions: Jahn-Teller Effect in Crystals and Molecules}}},\ \bibinfo
  {editor} {edited by\ \bibinfo {editor} {\bibfnamefont {M.~D.}\ \bibnamefont
  {Kaplan}}\ and\ \bibinfo {editor} {\bibfnamefont {G.~O.}\ \bibnamefont
  {Zimmerman}}}\ (\bibinfo  {publisher} {Springer Netherlands},\ \bibinfo
  {address} {Dordrecht},\ \bibinfo {year} {2001})\ pp.\ \bibinfo {pages}
  {343--347}\BibitemShut {NoStop}%
\bibitem [{\citenamefont {Manley}\ \emph {et~al.}(2011)\citenamefont {Manley},
  \citenamefont {Abernathy}, \citenamefont {Agladze},\ and\ \citenamefont
  {Sievers}}]{Manley:2011aa}%
  \BibitemOpen
  \bibfield  {author} {\bibinfo {author} {\bibfnamefont {M.~E.}\ \bibnamefont
  {Manley}}, \bibinfo {author} {\bibfnamefont {D.~L.}\ \bibnamefont
  {Abernathy}}, \bibinfo {author} {\bibfnamefont {N.~I.}\ \bibnamefont
  {Agladze}}, \ and\ \bibinfo {author} {\bibfnamefont {A.~J.}\ \bibnamefont
  {Sievers}},\ }\href@noop {} {\bibfield  {journal} {\bibinfo  {journal}
  {Scientific Reports}\ }\textbf {\bibinfo {volume} {1}},\ \bibinfo {pages}
  {Article number: 4} (\bibinfo {year} {2011})}\BibitemShut {NoStop}%
\bibitem [{\citenamefont {Kempa}\ \emph {et~al.}(2014)\citenamefont {Kempa},
  \citenamefont {Ondrejkovic}, \citenamefont {Bourges}, \citenamefont
  {Marton},\ and\ \citenamefont {Hlinka}}]{Kempa:2014aa}%
  \BibitemOpen
  \bibfield  {author} {\bibinfo {author} {\bibfnamefont {M.}~\bibnamefont
  {Kempa}}, \bibinfo {author} {\bibfnamefont {P.}~\bibnamefont {Ondrejkovic}},
  \bibinfo {author} {\bibfnamefont {P.}~\bibnamefont {Bourges}}, \bibinfo
  {author} {\bibfnamefont {P.}~\bibnamefont {Marton}}, \ and\ \bibinfo {author}
  {\bibfnamefont {J.}~\bibnamefont {Hlinka}},\ }\href {\doibase
  10.1103/PhysRevB.89.054308} {\bibfield  {journal} {\bibinfo  {journal}
  {Physical Review B - Condensed Matter and Materials Physics}\ }\textbf
  {\bibinfo {volume} {89}},\ \bibinfo {pages} {054308} (\bibinfo {year}
  {2014})}\BibitemShut {NoStop}%
\bibitem [{\citenamefont {Manley}\ \emph {et~al.}(2014)\citenamefont {Manley},
  \citenamefont {Jeffries}, \citenamefont {Lee}, \citenamefont {Butch},
  \citenamefont {Zabalegui},\ and\ \citenamefont {Abernathy}}]{Manley:2014aa}%
  \BibitemOpen
  \bibfield  {author} {\bibinfo {author} {\bibfnamefont {M.~E.}\ \bibnamefont
  {Manley}}, \bibinfo {author} {\bibfnamefont {J.~R.}\ \bibnamefont
  {Jeffries}}, \bibinfo {author} {\bibfnamefont {H.}~\bibnamefont {Lee}},
  \bibinfo {author} {\bibfnamefont {N.~P.}\ \bibnamefont {Butch}}, \bibinfo
  {author} {\bibfnamefont {A.}~\bibnamefont {Zabalegui}}, \ and\ \bibinfo
  {author} {\bibfnamefont {D.~L.}\ \bibnamefont {Abernathy}},\ }\href@noop {}
  {\bibfield  {journal} {\bibinfo  {journal} {Physical Review B}\ }\textbf
  {\bibinfo {volume} {89}},\ \bibinfo {pages} {224106} (\bibinfo {year}
  {2014})}\BibitemShut {NoStop}%
\bibitem [{\citenamefont {Plimpton}(1995)}]{Plimpton:1995aa}%
  \BibitemOpen
  \bibfield  {author} {\bibinfo {author} {\bibfnamefont {S.}~\bibnamefont
  {Plimpton}},\ }\href@noop {} {\bibfield  {journal} {\bibinfo  {journal}
  {Journal of Computational Physics}\ }\textbf {\bibinfo {volume} {117}},\
  \bibinfo {pages} {1} (\bibinfo {year} {1995})}\BibitemShut {NoStop}%
\bibitem [{Note1()}]{Note1}%
  \BibitemOpen
  \bibinfo {note} {We observe that PBCs with $N_c=10$ appears a safe choice to
  inspect energy localization on length scales of the order of half/one unit
  cell.}\BibitemShut {Stop}%
\bibitem [{\citenamefont {Ruffa}(1963)}]{Ruffa:1963aa}%
  \BibitemOpen
  \bibfield  {author} {\bibinfo {author} {\bibfnamefont {A.~R.}\ \bibnamefont
  {Ruffa}},\ }\href@noop {} {\bibfield  {journal} {\bibinfo  {journal} {Phys.
  Rev.}\ }\textbf {\bibinfo {volume} {130}},\ \bibinfo {pages} {1412} (\bibinfo
  {year} {1963})}\BibitemShut {NoStop}%
\bibitem [{\citenamefont {Rapp}\ and\ \citenamefont
  {Merchant}(1973)}]{Rapp:1973aa}%
  \BibitemOpen
  \bibfield  {author} {\bibinfo {author} {\bibfnamefont {J.~E.}\ \bibnamefont
  {Rapp}}\ and\ \bibinfo {author} {\bibfnamefont {H.~D.}\ \bibnamefont
  {Merchant}},\ }\href@noop {} {\bibfield  {journal} {\bibinfo  {journal}
  {Journal of Applied Physics}\ }\textbf {\bibinfo {volume} {44}},\ \bibinfo
  {pages} {3919} (\bibinfo {year} {1973})}\BibitemShut {NoStop}%
\bibitem [{\citenamefont {Sangster}(1976)}]{Sangster:1976aa}%
  \BibitemOpen
  \bibfield  {author} {\bibinfo {author} {\bibfnamefont {M.~J.~L.}\
  \bibnamefont {Sangster}},\ }\href@noop {} {\bibfield  {journal} {\bibinfo
  {journal} {Solid State Communications}\ }\textbf {\bibinfo {volume} {18}},\
  \bibinfo {pages} {67} (\bibinfo {year} {1976})}\BibitemShut {NoStop}%
\bibitem [{\citenamefont {Jain}\ \emph {et~al.}(1976)\citenamefont {Jain},
  \citenamefont {Shanker},\ and\ \citenamefont {Khandelwal}}]{Jain:1976aa}%
  \BibitemOpen
  \bibfield  {author} {\bibinfo {author} {\bibfnamefont {J.~K.}\ \bibnamefont
  {Jain}}, \bibinfo {author} {\bibfnamefont {J.}~\bibnamefont {Shanker}}, \
  and\ \bibinfo {author} {\bibfnamefont {D.~P.}\ \bibnamefont {Khandelwal}},\
  }\href@noop {} {\bibfield  {journal} {\bibinfo  {journal} {Phys. Rev. B}\
  }\textbf {\bibinfo {volume} {13}},\ \bibinfo {pages} {2692} (\bibinfo {year}
  {1976})}\BibitemShut {NoStop}%
\bibitem [{\citenamefont {Sangster}\ \emph {et~al.}(1978)\citenamefont
  {Sangster}, \citenamefont {Schr\"oder},\ and\ \citenamefont
  {Atwood}}]{Sangster:1978aa}%
  \BibitemOpen
  \bibfield  {author} {\bibinfo {author} {\bibfnamefont {M.~J.~L.}\
  \bibnamefont {Sangster}}, \bibinfo {author} {\bibfnamefont {U.}~\bibnamefont
  {Schr\"oder}}, \ and\ \bibinfo {author} {\bibfnamefont {R.~M.}\ \bibnamefont
  {Atwood}},\ }\href@noop {} {\bibfield  {journal} {\bibinfo  {journal} {J.
  Phys. C: Solid State Phys.}\ }\textbf {\bibinfo {volume} {11}},\ \bibinfo
  {pages} {1523} (\bibinfo {year} {1978})}\BibitemShut {NoStop}%
\bibitem [{\citenamefont {Shanker}\ \emph {et~al.}(1984)\citenamefont
  {Shanker}, \citenamefont {Narain},\ and\ \citenamefont
  {Rajauria}}]{Shanker:1984aa}%
  \BibitemOpen
  \bibfield  {author} {\bibinfo {author} {\bibfnamefont {J.}~\bibnamefont
  {Shanker}}, \bibinfo {author} {\bibfnamefont {J.}~\bibnamefont {Narain}}, \
  and\ \bibinfo {author} {\bibfnamefont {A.~K.}\ \bibnamefont {Rajauria}},\
  }\href@noop {} {\bibfield  {journal} {\bibinfo  {journal} {Physica Status
  Solidi B}\ }\textbf {\bibinfo {volume} {122}},\ \bibinfo {pages} {435}
  (\bibinfo {year} {1984})}\BibitemShut {NoStop}%
\bibitem [{\citenamefont {Shanker}\ and\ \citenamefont
  {Agrawal}(1984)}]{Shanker:1984ab}%
  \BibitemOpen
  \bibfield  {author} {\bibinfo {author} {\bibfnamefont {J.}~\bibnamefont
  {Shanker}}\ and\ \bibinfo {author} {\bibfnamefont {G.~G.}\ \bibnamefont
  {Agrawal}},\ }\href@noop {} {\bibfield  {journal} {\bibinfo  {journal}
  {Physica Status Solidi B}\ }\textbf {\bibinfo {volume} {123}},\ \bibinfo
  {pages} {11} (\bibinfo {year} {1984})}\BibitemShut {NoStop}%
\bibitem [{\citenamefont {Kumar}\ \emph {et~al.}(1986)\citenamefont {Kumar},
  \citenamefont {Kaur},\ and\ \citenamefont {Shanker}}]{Kumar:1986aa}%
  \BibitemOpen
  \bibfield  {author} {\bibinfo {author} {\bibfnamefont {M.}~\bibnamefont
  {Kumar}}, \bibinfo {author} {\bibfnamefont {A.~J.}\ \bibnamefont {Kaur}}, \
  and\ \bibinfo {author} {\bibfnamefont {J.}~\bibnamefont {Shanker}},\
  }\href@noop {} {\bibfield  {journal} {\bibinfo  {journal} {Journal of
  Chemical Physics}\ }\textbf {\bibinfo {volume} {84}},\ \bibinfo {pages}
  {5735} (\bibinfo {year} {1986})}\BibitemShut {NoStop}%
\bibitem [{\citenamefont {Sherry}\ and\ \citenamefont
  {Kumar}(1991)}]{Sherry:1991aa}%
  \BibitemOpen
  \bibfield  {author} {\bibinfo {author} {\bibfnamefont {A.~M.}\ \bibnamefont
  {Sherry}}\ and\ \bibinfo {author} {\bibfnamefont {M.}~\bibnamefont {Kumar}},\
  }\href@noop {} {\bibfield  {journal} {\bibinfo  {journal} {J. Phys. Chem.
  Solids}\ }\textbf {\bibinfo {volume} {52}},\ \bibinfo {pages} {1145}
  (\bibinfo {year} {1991})}\BibitemShut {NoStop}%
\bibitem [{\citenamefont {Sunil}\ and\ \citenamefont
  {Sharma}(2012)}]{Sumil:2012aa}%
  \BibitemOpen
  \bibfield  {author} {\bibinfo {author} {\bibfnamefont {K.}~\bibnamefont
  {Sunil}}\ and\ \bibinfo {author} {\bibfnamefont {B.~S.}\ \bibnamefont
  {Sharma}},\ }\href@noop {} {\bibfield  {journal} {\bibinfo  {journal} {Indian
  Journal of Pure and Applied Physics}\ }\textbf {\bibinfo {volume} {50}},\
  \bibinfo {pages} {387} (\bibinfo {year} {2012})}\BibitemShut {NoStop}%
\bibitem [{\citenamefont {Ewald}(1921)}]{Ewald:1921aa}%
  \BibitemOpen
  \bibfield  {author} {\bibinfo {author} {\bibfnamefont {P.~P.}\ \bibnamefont
  {Ewald}},\ }\href@noop {} {\bibfield  {journal} {\bibinfo  {journal} {Annalen
  der Physik}\ }\textbf {\bibinfo {volume} {369}},\ \bibinfo {pages} {253}
  (\bibinfo {year} {1921})}\BibitemShut {NoStop}%
\bibitem [{\citenamefont {Muller}(1936)}]{Muller:1936aa}%
  \BibitemOpen
  \bibfield  {author} {\bibinfo {author} {\bibfnamefont {A.}~\bibnamefont
  {Muller}},\ }\href@noop {} {\bibfield  {journal} {\bibinfo  {journal} {proc.
  Roy. Soc. (London)}\ }\textbf {\bibinfo {volume} {A154}},\ \bibinfo {pages}
  {624} (\bibinfo {year} {1936})}\BibitemShut {NoStop}%
\bibitem [{\citenamefont {Kirkwood}(1932)}]{Kirkwood:1932aa}%
  \BibitemOpen
  \bibfield  {author} {\bibinfo {author} {\bibfnamefont {J.~G.}\ \bibnamefont
  {Kirkwood}},\ }\href@noop {} {\bibfield  {journal} {\bibinfo  {journal} {Z.
  Phys. (Leipzig)}\ }\textbf {\bibinfo {volume} {33}},\ \bibinfo {pages} {57}
  (\bibinfo {year} {1932})}\BibitemShut {NoStop}%
\bibitem [{\citenamefont {Harding}(1991)}]{Harding:1991aa}%
  \BibitemOpen
  \bibfield  {author} {\bibinfo {author} {\bibfnamefont {J.~H.}\ \bibnamefont
  {Harding}},\ }\enquote {\bibinfo {title} {Computer simulation in materials
  science},}\ \ (\bibinfo  {publisher} {Kluwer Academic Publishers},\ \bibinfo
  {address} {Dordrecht},\ \bibinfo {year} {1991})\ Chap.\ \bibinfo {chapter}
  {Interatomic potentials: a user guide}\BibitemShut {NoStop}%
\bibitem [{\citenamefont {Nos\'e}(1984)}]{Nose:1984aa}%
  \BibitemOpen
  \bibfield  {author} {\bibinfo {author} {\bibfnamefont {S.}~\bibnamefont
  {Nos\'e}},\ }\href@noop {} {\bibfield  {journal} {\bibinfo  {journal}
  {Molecular Physics}\ }\textbf {\bibinfo {volume} {52}},\ \bibinfo {pages}
  {255} (\bibinfo {year} {1984})}\BibitemShut {NoStop}%
\bibitem [{\citenamefont {Hoover}(1985)}]{Hoover:1985aa}%
  \BibitemOpen
  \bibfield  {author} {\bibinfo {author} {\bibfnamefont {W.~G.}\ \bibnamefont
  {Hoover}},\ }\href@noop {} {\bibfield  {journal} {\bibinfo  {journal} {Phys.
  Rev. A}\ }\textbf {\bibinfo {volume} {31}},\ \bibinfo {pages} {1695}
  (\bibinfo {year} {1985})}\BibitemShut {NoStop}%
\bibitem [{Note2()}]{Note2}%
  \BibitemOpen
  \bibinfo {note} {The use of an NVT dynamics for production runs does not
  appear to make sense in this study. In fact, thermostats are in principle
  nothing but {\protect \em smart sampling} techniques, designed to produce
  time series sampled from the canonical measure. However, there is absolutely
  no guarantee that the actual trajectories (i.e. the actual {\protect \em
  dynamics}) make any physical sense. In particular, all vibrational coherences
  are either (artificially) damped or completely destroyed, depending on the
  value of the relaxation time scale chosen for the specific thermostat. In
  practice, it is preferable to switch off the thermostat once the system has
  reached thermal equilibrium, so that no artificial noise is left to fiddle
  with the vibrational coherences that might emerge in specific frequency
  regions.}\BibitemShut {Stop}%
\bibitem [{\citenamefont {Gale}\ and\ \citenamefont
  {Rohl}(2003)}]{Gale:2003aa}%
  \BibitemOpen
  \bibfield  {author} {\bibinfo {author} {\bibfnamefont {J.~D.}\ \bibnamefont
  {Gale}}\ and\ \bibinfo {author} {\bibfnamefont {A.~L.}\ \bibnamefont
  {Rohl}},\ }\href {\doibase 10.1080/0892702031000104887} {\bibfield  {journal}
  {\bibinfo  {journal} {Molecular Simulation}\ }\textbf {\bibinfo {volume}
  {29}},\ \bibinfo {pages} {291} (\bibinfo {year} {2003})}\BibitemShut
  {NoStop}%
\bibitem [{\citenamefont {Forinash}\ and\ \citenamefont
  {Lang}(1998)}]{Forinash:1998aa}%
  \BibitemOpen
  \bibfield  {author} {\bibinfo {author} {\bibfnamefont {K.}~\bibnamefont
  {Forinash}}\ and\ \bibinfo {author} {\bibfnamefont {W.}~\bibnamefont
  {Lang}},\ }\href@noop {} {\bibfield  {journal} {\bibinfo  {journal} {Physica
  D: Nonlinear Phenomena}\ }\textbf {\bibinfo {volume} {123}},\ \bibinfo
  {pages} {437} (\bibinfo {year} {1998})}\BibitemShut {NoStop}%
\bibitem [{\citenamefont {Gabor}(1946)}]{Gabor:1946aa}%
  \BibitemOpen
  \bibfield  {author} {\bibinfo {author} {\bibfnamefont {D.}~\bibnamefont
  {Gabor}},\ }\href@noop {} {\bibfield  {journal} {\bibinfo  {journal} {J.
  Inst. Elec. Eng. (London)}\ }\textbf {\bibinfo {volume} {93 (III)}},\
  \bibinfo {pages} {429} (\bibinfo {year} {1946})}\BibitemShut {NoStop}%
\bibitem [{\citenamefont {Cretegny}\ \emph
  {et~al.}(1998{\natexlab{a}})\citenamefont {Cretegny}, \citenamefont {Livi},\
  and\ \citenamefont {Spicci}}]{Cretegny:1998aa}%
  \BibitemOpen
  \bibfield  {author} {\bibinfo {author} {\bibfnamefont {T.}~\bibnamefont
  {Cretegny}}, \bibinfo {author} {\bibfnamefont {R.}~\bibnamefont {Livi}}, \
  and\ \bibinfo {author} {\bibfnamefont {M.}~\bibnamefont {Spicci}},\ }\href
  {\doibase 10.1016/S0167-2789(98)00080-3} {\bibfield  {journal} {\bibinfo
  {journal} {Physica D: Nonlinear Phenomena}\ }\textbf {\bibinfo {volume}
  {119}},\ \bibinfo {pages} {88} (\bibinfo {year}
  {1998}{\natexlab{a}})}\BibitemShut {NoStop}%
\bibitem [{Note3()}]{Note3}%
  \BibitemOpen
  \bibinfo {note} {In this work we implicitly refer to the gap spectral region
  when we mention the excitation of a burst.}\BibitemShut {Stop}%
\bibitem [{\citenamefont {Khadeeva}\ and\ \citenamefont
  {Dmitriev}(2011)}]{Khadeeva:2011aa}%
  \BibitemOpen
  \bibfield  {author} {\bibinfo {author} {\bibfnamefont {L.~Z.}\ \bibnamefont
  {Khadeeva}}\ and\ \bibinfo {author} {\bibfnamefont {S.~V.}\ \bibnamefont
  {Dmitriev}},\ }\href@noop {} {\bibfield  {journal} {\bibinfo  {journal}
  {Physical Review B}\ }\textbf {\bibinfo {volume} {84}},\ \bibinfo {pages}
  {144304} (\bibinfo {year} {2011})}\BibitemShut {NoStop}%
\bibitem [{\citenamefont {Flach}\ \emph {et~al.}(1997)\citenamefont {Flach},
  \citenamefont {Kladko},\ and\ \citenamefont {MacKay}}]{Flach:1997qy}%
  \BibitemOpen
  \bibfield  {author} {\bibinfo {author} {\bibfnamefont {S.}~\bibnamefont
  {Flach}}, \bibinfo {author} {\bibfnamefont {K.}~\bibnamefont {Kladko}}, \
  and\ \bibinfo {author} {\bibfnamefont {R.~S.}\ \bibnamefont {MacKay}},\
  }\href@noop {} {\bibfield  {journal} {\bibinfo  {journal} {Physical Review
  Letters}\ }\textbf {\bibinfo {volume} {78}},\ \bibinfo {pages} {1207}
  (\bibinfo {year} {1997})}\BibitemShut {NoStop}%
\bibitem [{\citenamefont {Kastner}(2004)}]{Kastner:2004fk}%
  \BibitemOpen
  \bibfield  {author} {\bibinfo {author} {\bibfnamefont {M.}~\bibnamefont
  {Kastner}},\ }\href@noop {} {\bibfield  {journal} {\bibinfo  {journal}
  {Nonlinearity}\ }\textbf {\bibinfo {volume} {17}},\ \bibinfo {pages} {1923}
  (\bibinfo {year} {2004})}\BibitemShut {NoStop}%
\bibitem [{\citenamefont {Piazza}\ \emph {et~al.}(2003)\citenamefont {Piazza},
  \citenamefont {Lepri},\ and\ \citenamefont {Livi}}]{Piazza:2003aa}%
  \BibitemOpen
  \bibfield  {author} {\bibinfo {author} {\bibfnamefont {F.}~\bibnamefont
  {Piazza}}, \bibinfo {author} {\bibfnamefont {S.}~\bibnamefont {Lepri}}, \
  and\ \bibinfo {author} {\bibfnamefont {R.}~\bibnamefont {Livi}},\ }\href@noop
  {} {\bibfield  {journal} {\bibinfo  {journal} {Chaos}\ }\textbf {\bibinfo
  {volume} {13}},\ \bibinfo {pages} {637} (\bibinfo {year} {2003})}\BibitemShut
  {NoStop}%
\bibitem [{\citenamefont {Cretegny}\ \emph
  {et~al.}(1998{\natexlab{b}})\citenamefont {Cretegny}, \citenamefont
  {Dauxois}, \citenamefont {Ruffo},\ and\ \citenamefont
  {Torcini}}]{Cretegny:1998ab}%
  \BibitemOpen
  \bibfield  {author} {\bibinfo {author} {\bibfnamefont {T.}~\bibnamefont
  {Cretegny}}, \bibinfo {author} {\bibfnamefont {T.}~\bibnamefont {Dauxois}},
  \bibinfo {author} {\bibfnamefont {S.}~\bibnamefont {Ruffo}}, \ and\ \bibinfo
  {author} {\bibfnamefont {A.}~\bibnamefont {Torcini}},\ }\href@noop {}
  {\bibfield  {journal} {\bibinfo  {journal} {Physica D: Nonlinear Phenomena}\
  }\textbf {\bibinfo {volume} {121}},\ \bibinfo {pages} {109} (\bibinfo {year}
  {1998}{\natexlab{b}})}\BibitemShut {NoStop}%
\bibitem [{\citenamefont {Kosevich}\ and\ \citenamefont
  {Lepri}(2000)}]{Kosevich:2000aa}%
  \BibitemOpen
  \bibfield  {author} {\bibinfo {author} {\bibfnamefont {Y.~A.}\ \bibnamefont
  {Kosevich}}\ and\ \bibinfo {author} {\bibfnamefont {S.}~\bibnamefont
  {Lepri}},\ }\href@noop {} {\bibfield  {journal} {\bibinfo  {journal} {Phys.
  Rev. B}\ }\textbf {\bibinfo {volume} {61}},\ \bibinfo {pages} {299} (\bibinfo
  {year} {2000})}\BibitemShut {NoStop}%
\bibitem [{\citenamefont {Wrubel}\ \emph {et~al.}(2005)\citenamefont {Wrubel},
  \citenamefont {Sato},\ and\ \citenamefont {Sievers}}]{Wrubel:2005aa}%
  \BibitemOpen
  \bibfield  {author} {\bibinfo {author} {\bibfnamefont {J.~P.}\ \bibnamefont
  {Wrubel}}, \bibinfo {author} {\bibfnamefont {M.}~\bibnamefont {Sato}}, \ and\
  \bibinfo {author} {\bibfnamefont {A.~J.}\ \bibnamefont {Sievers}},\
  }\href@noop {} {\bibfield  {journal} {\bibinfo  {journal} {Phys. Rev. Lett.}\
  }\textbf {\bibinfo {volume} {95}},\ \bibinfo {pages} {264101} (\bibinfo
  {year} {2005})}\BibitemShut {NoStop}%
\bibitem [{\citenamefont {Sato}\ and\ \citenamefont
  {Sievers}(2005)}]{Sato:2005aa}%
  \BibitemOpen
  \bibfield  {author} {\bibinfo {author} {\bibfnamefont {M.}~\bibnamefont
  {Sato}}\ and\ \bibinfo {author} {\bibfnamefont {A.~J.}\ \bibnamefont
  {Sievers}},\ }\href@noop {} {\bibfield  {journal} {\bibinfo  {journal} {Phys.
  Rev. B}\ }\textbf {\bibinfo {volume} {71}},\ \bibinfo {pages} {214306}
  (\bibinfo {year} {2005})}\BibitemShut {NoStop}%
\bibitem [{\citenamefont {Pouyandeh}\ \emph {et~al.}(2017)\citenamefont
  {Pouyandeh}, \citenamefont {Iubini}, \citenamefont {Jurinovich},
  \citenamefont {Omar}, \citenamefont {Mennucci},\ and\ \citenamefont
  {Piazza}}]{Pouyandeh:2017aa}%
  \BibitemOpen
  \bibfield  {author} {\bibinfo {author} {\bibfnamefont {S.}~\bibnamefont
  {Pouyandeh}}, \bibinfo {author} {\bibfnamefont {S.}~\bibnamefont {Iubini}},
  \bibinfo {author} {\bibfnamefont {S.}~\bibnamefont {Jurinovich}}, \bibinfo
  {author} {\bibfnamefont {Y.}~\bibnamefont {Omar}}, \bibinfo {author}
  {\bibfnamefont {B.}~\bibnamefont {Mennucci}}, \ and\ \bibinfo {author}
  {\bibfnamefont {F.}~\bibnamefont {Piazza}},\ }\href@noop {} {\bibfield
  {journal} {\bibinfo  {journal} {Physical Biology}\ }\textbf {\bibinfo
  {volume} {14}},\ \bibinfo {pages} {066001} (\bibinfo {year}
  {2017})}\BibitemShut {NoStop}%
\bibitem [{\citenamefont {Viani}\ \emph {et~al.}(2014)\citenamefont {Viani},
  \citenamefont {Corbella}, \citenamefont {Curutchet}, \citenamefont
  {O'Reilly}, \citenamefont {Olaya-Castro},\ and\ \citenamefont
  {Mennucci}}]{Viani:2014kl}%
  \BibitemOpen
  \bibfield  {author} {\bibinfo {author} {\bibfnamefont {L.}~\bibnamefont
  {Viani}}, \bibinfo {author} {\bibfnamefont {M.}~\bibnamefont {Corbella}},
  \bibinfo {author} {\bibfnamefont {C.}~\bibnamefont {Curutchet}}, \bibinfo
  {author} {\bibfnamefont {E.~J.}\ \bibnamefont {O'Reilly}}, \bibinfo {author}
  {\bibfnamefont {A.}~\bibnamefont {Olaya-Castro}}, \ and\ \bibinfo {author}
  {\bibfnamefont {B.}~\bibnamefont {Mennucci}},\ }\href@noop {} {\bibfield
  {journal} {\bibinfo  {journal} {Physical Chemistry Chemical Physics}\
  }\textbf {\bibinfo {volume} {16}},\ \bibinfo {pages} {16302} (\bibinfo {year}
  {2014})}\BibitemShut {NoStop}%
\bibitem [{\citenamefont {Wu}\ \emph {et~al.}(2010)\citenamefont {Wu},
  \citenamefont {Liu}, \citenamefont {Shen}, \citenamefont {Cao},\ and\
  \citenamefont {Silbey}}]{Wu:2010aa}%
  \BibitemOpen
  \bibfield  {author} {\bibinfo {author} {\bibfnamefont {J.}~\bibnamefont
  {Wu}}, \bibinfo {author} {\bibfnamefont {F.}~\bibnamefont {Liu}}, \bibinfo
  {author} {\bibfnamefont {Y.}~\bibnamefont {Shen}}, \bibinfo {author}
  {\bibfnamefont {J.}~\bibnamefont {Cao}}, \ and\ \bibinfo {author}
  {\bibfnamefont {R.~J.}\ \bibnamefont {Silbey}},\ }\href@noop {} {\bibfield
  {journal} {\bibinfo  {journal} {New Journal of Physics}\ }\textbf {\bibinfo
  {volume} {12}},\ \bibinfo {pages} {105012} (\bibinfo {year}
  {2010})}\BibitemShut {NoStop}%
\bibitem [{\citenamefont {Yu}\ and\ \citenamefont {Leitner}(2003)}]{Yu:2003fk}%
  \BibitemOpen
  \bibfield  {author} {\bibinfo {author} {\bibfnamefont {X.}~\bibnamefont
  {Yu}}\ and\ \bibinfo {author} {\bibfnamefont {D.}~\bibnamefont {Leitner}},\
  }\href@noop {} {\bibfield  {journal} {\bibinfo  {journal} {Journal of
  Physical Chemistry B}\ }\textbf {\bibinfo {volume} {107}},\ \bibinfo {pages}
  {1698} (\bibinfo {year} {2003})}\BibitemShut {NoStop}%
\bibitem [{\citenamefont {Leitner}(2008)}]{Leitner:2008gl}%
  \BibitemOpen
  \bibfield  {author} {\bibinfo {author} {\bibfnamefont {D.~M.}\ \bibnamefont
  {Leitner}},\ }\href@noop {} {\bibfield  {journal} {\bibinfo  {journal}
  {Annual Review of Physical Chemistry}\ }\textbf {\bibinfo {volume} {59}},\
  \bibinfo {pages} {233} (\bibinfo {year} {2008})}\BibitemShut {NoStop}%
\bibitem [{\citenamefont {Sharp}\ and\ \citenamefont
  {Skinner}(2006)}]{Sharp:2006aa}%
  \BibitemOpen
  \bibfield  {author} {\bibinfo {author} {\bibfnamefont {K.}~\bibnamefont
  {Sharp}}\ and\ \bibinfo {author} {\bibfnamefont {J.~J.}\ \bibnamefont
  {Skinner}},\ }\href {\doibase 10.1002/prot.21146} {\bibfield  {journal}
  {\bibinfo  {journal} {Proteins: Structure, Function, and Bioinformatics}\
  }\textbf {\bibinfo {volume} {65}},\ \bibinfo {pages} {347} (\bibinfo {year}
  {2006})}\BibitemShut {NoStop}%
\bibitem [{\citenamefont {Piazza}\ and\ \citenamefont
  {Sanejouand}(2009)}]{Piazza:2009aa}%
  \BibitemOpen
  \bibfield  {author} {\bibinfo {author} {\bibfnamefont {F.}~\bibnamefont
  {Piazza}}\ and\ \bibinfo {author} {\bibfnamefont {Y.-H.}\ \bibnamefont
  {Sanejouand}},\ }\href {http://stacks.iop.org/0295-5075/88/i=6/a=68001}
  {\bibfield  {journal} {\bibinfo  {journal} {EPL (Europhysics Letters)}\
  }\textbf {\bibinfo {volume} {88}},\ \bibinfo {pages} {68001} (\bibinfo {year}
  {2009})}\BibitemShut {NoStop}%
\bibitem [{\citenamefont {Tsai}\ and\ \citenamefont
  {Nussinov}(2014)}]{Tsai:2014aa}%
  \BibitemOpen
  \bibfield  {author} {\bibinfo {author} {\bibfnamefont {C.-J.}\ \bibnamefont
  {Tsai}}\ and\ \bibinfo {author} {\bibfnamefont {R.}~\bibnamefont
  {Nussinov}},\ }\href@noop {} {\bibfield  {journal} {\bibinfo  {journal} {PLOS
  Computational Biology}\ }\textbf {\bibinfo {volume} {10}},\ \bibinfo {pages}
  {e1003394} (\bibinfo {year} {2014})}\BibitemShut {NoStop}%
\bibitem [{\citenamefont {Allain}\ \emph {et~al.}(2014)\citenamefont {Allain},
  \citenamefont {{Chauvot de Beauchene}}, \citenamefont {Langenfeld},
  \citenamefont {Guarracino}, \citenamefont {Laine},\ and\ \citenamefont
  {Tchertanov}}]{Allain:2014aa}%
  \BibitemOpen
  \bibfield  {author} {\bibinfo {author} {\bibfnamefont {A.}~\bibnamefont
  {Allain}}, \bibinfo {author} {\bibfnamefont {I.}~\bibnamefont {{Chauvot de
  Beauchene}}}, \bibinfo {author} {\bibfnamefont {F.}~\bibnamefont
  {Langenfeld}}, \bibinfo {author} {\bibfnamefont {Y.}~\bibnamefont
  {Guarracino}}, \bibinfo {author} {\bibfnamefont {E.}~\bibnamefont {Laine}}, \
  and\ \bibinfo {author} {\bibfnamefont {L.}~\bibnamefont {Tchertanov}},\
  }\href {\doibase 10.1039/C4FD00024B} {\bibfield  {journal} {\bibinfo
  {journal} {Faraday Discuss.}\ }\textbf {\bibinfo {volume} {169}},\ \bibinfo
  {pages} {303} (\bibinfo {year} {2014})}\BibitemShut {NoStop}%
\end{thebibliography}
%

%


\end{document}